\documentclass[aip,
 amsmath,amssymb,
 preprint,%
a4paper
]{revtex4-1}

\usepackage{graphicx}
\usepackage{dcolumn}
\usepackage{bm}

\usepackage[utf8]{inputenc}
\usepackage[T1]{fontenc}
\usepackage{mathptmx}
\usepackage{etoolbox}
\usepackage[table]{xcolor}
\usepackage[colorlinks=true,linkcolor=blue,citecolor=blue,urlcolor=blue,allbordercolors=white]{hyperref}

\begin{document}

\title{\large A Multiscale Sensorimotor Model of Experience-Dependent Behavior in a Minimal Organism}

\author{Mar\'ia Sol Vidal-Saez}
\affiliation{Department of Medicine and Life Sciences, Universitat Pompeu Fabra, Dr Aiguader 88, 08003, Barcelona, Spain}

\author{Oscar Vilarroya}
\affiliation{Department of Psychiatry and Legal Medicine, Universitat Aut\`onoma de Barcelona, 08193, Cerdanyola del Vallès, Spain}
\affiliation{Hospital del Mar Medical Research Institute (IMIM), Dr Aiguader 88, 08003, Barcelona, Spain}

\author{Jordi Garcia-Ojalvo}
\affiliation{Department of Medicine and Life Sciences, Universitat Pompeu Fabra, Dr Aiguader 88, 08003, Barcelona, Spain}

\begin{abstract}
To survive in ever-changing environments, living organisms need to continuously combine the ongoing external inputs they receive, representing present conditions, with their dynamical internal state, which includes influences of past experiences.
It is still unclear in general, however, (i) how this happens at the molecular and cellular levels, and (ii) how the corresponding molecular and cellular processes are integrated with the behavioral responses of the organism.
Here we address these issues by modeling mathematically a particular behavioral paradigm in a minimal model organism, namely chemotaxis in the nematode \textit{C. elegans}.
Specifically, we use a long-standing collection of elegant experiments on salt chemotaxis in this animal, in which the migration direction varies depending on its previous experience.
Our model integrates the molecular, cellular and organismal levels to reproduce the experimentally observed experience-dependent behavior.
The model proposes specific molecular mechanisms for the encoding of current conditions and past experiences in key neurons associated with this response, predicting the behavior of various mutants associated with those molecular circuits.
\end{abstract}





\maketitle

\section{Introduction}
\label{intro}

Every organism makes decisions many times a day, and innumerable times in their lifetime.
These decision-making events are crucial for survival.
This is the case, for example, when animals navigate their complex and dynamic environment in search of food, a mate or a breeding site, or to escape from predators.
In order to make decisions, animals usually evaluate signals from their current external environment and integrate them with their past experiences (which are encoded in their internal state) \cite{fukushi2004navigation, hemingway2020state,luo2021sex}. 
This results in experience-dependent behaviors, where the individual's past experiences (or context) affect their present actions. One of the most important topics of current research is to identify the molecular and cellular mechanisms that underpin these adaptive behaviors.

The nematode \textit{Caenorhabditis elegans} is an ideal organism in which to explore the neural basis of experience-dependent behaviors. \textit{C. elegans} has been an important model system for biological research over the years in a variety of fields including genomics, cell biology, developmental biology, and neuroscience \cite{brenner1974genetics, byerly1976life, durbin1987studies, bono2005neuronal, sengupta2009caenorhabditis}. Importantly for neuroscience, in particular, the \textit{C. elegans} connectome is by far the most complete to date, comprising 302 neurons and over 7000 connections for the hermaphrodite \cite{white1986structure}.
Despite its relatively simple nervous system, \textit{C. elegans} shows experience-dependent behavioral plasticity in response to a variety of environmental cues, such as odor \cite{colbert1995odorant}, taste \cite{saeki2001plasticity}, and temperature \cite{mori1995neural}, among others. 
 
\begin{figure}[htb]
\centerline{%
 \includegraphics[width=0.5\textwidth]{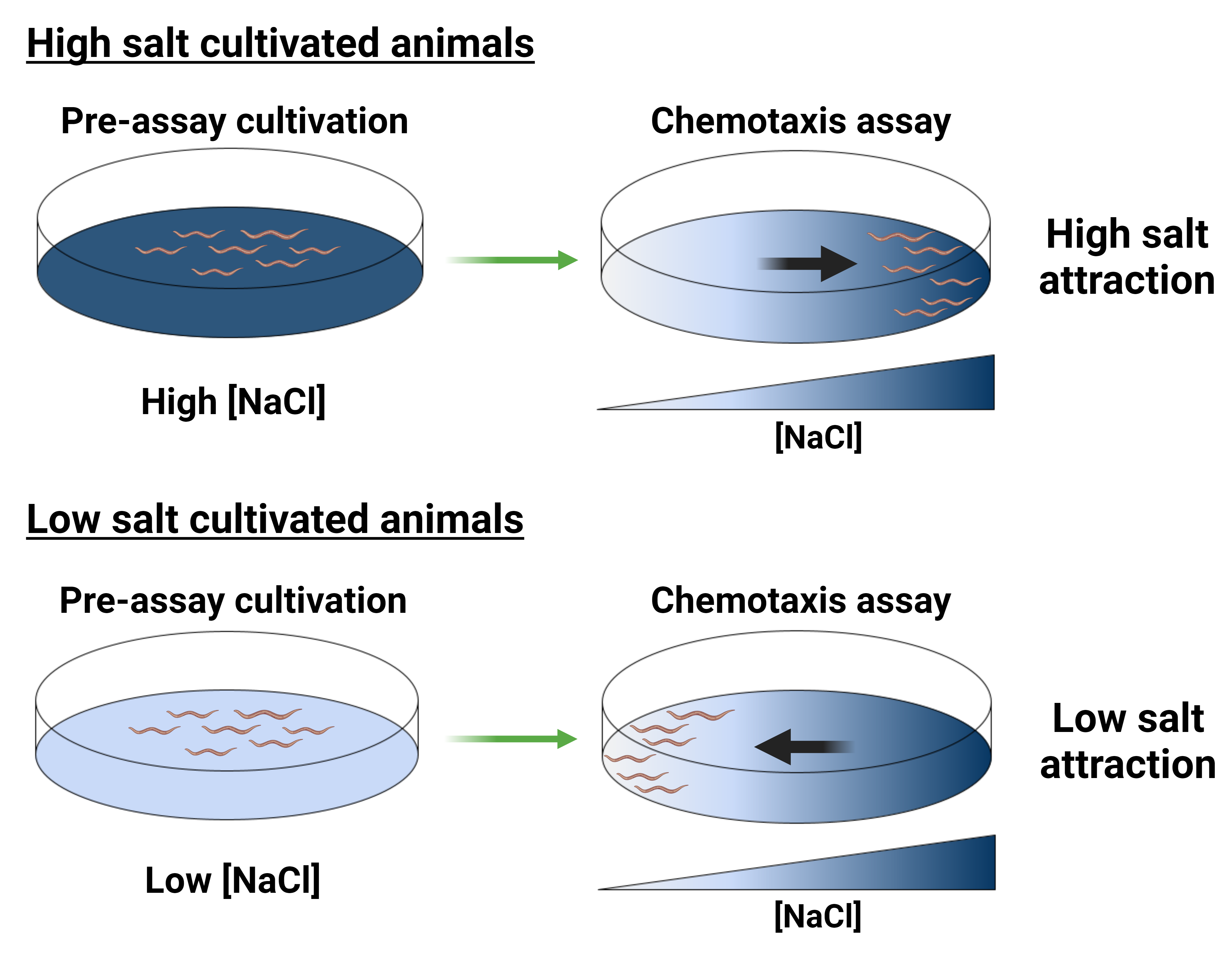}
}
\caption{\textbf{Experience-dependent salt chemotaxis in \textit{C. elegans}}. Worms are cultivated at either a high (top) or low (bottom) concentration of NaCl, then transferred to a plate with a salt gradient. High (low) salt cultivated animals are attracted to high (low) salt concentrations.}
\label{fig:1}  
\end{figure}

A series of elegant experiments performed over the years by the group of Prof. Yuichi Iino at the University of Tokyo  \cite{kunitomo2013concentration,ohno2017dynamics, sato2021glutamate, hiroki2022molecular} have demonstrated that chemotaxis of \textit{C. elegans} to sodium chloride (NaCl) is an experience-dependent behavior.
Specifically, animals migrate up or down salt gradients towards the concentration that they have previously experienced during cultivation \cite{kunitomo2013concentration}.
In other words, if a worm cultivated at a high NaCl concentration is set at a lower concentration in the middle of a salt gradient, it crawls up the gradient (top panel of Fig. \ref{fig:1}).
Conversely, if the worm was cultivated at a low NaCl concentration, it crawls down the gradient (bottom panel of Fig. \ref{fig:1}).

In this paper we present a mechanistic model, strongly grounded in multiple experimental observations at different levels, that captures the above-described experience-dependent chemotaxis to NaCl exhibited by \textit{C. elegans}.
The model includes within an integrated framework the sensory components of the process, at both the molecular and cellular levels, and the motor features of the behavior observed, at the organismal level.
The model is biologically grounded, including all necessary and sufficient components identified experimentally, without introducing artificial regulatory mechanisms. At the same time it is not too overly detailed, avoiding unnecessary molecular details that could lead to over-parameterization.
One of the highlights of the model is that within a single framework we can account for the two context-dependent behaviors, i.e. high- and low-salt attraction, according to the worm's previous experience, with the same circuit and without changing parameter values. 

\section{Neuronal determinants of experience-dependent chemotaxis} 

\label{sec:neural}
The neural and molecular mechanisms that underlie the experience-dependent behavior schematized in Fig.~\ref{fig:1} have been studied experimentally in a series of landmark articles by the Iino group \cite{kunitomo2013concentration, ohno2017dynamics, sato2021glutamate, hiroki2022molecular}.
These studies have revealed several key players at both the cellular and molecular levels.
At the cellular level, the amphid sensory neuron right (ASER) has been found to be necessary and sufficient for the behavior to occur.
Specifically, functional inhibition or genetic ablation of ASER completely impairs the plasticity of salt chemotaxis, whereas disruption of the activity of other chemosensory neurons has no substantial effect \cite{kunitomo2013concentration}.
This demonstrates that sensory input to ASER is critical for this adaptive behavior.
Taking this into account, in our model we limit the sensory aspects of the system to this neuron. 

\begin{figure}[htb]
\centerline{
\includegraphics[width=0.95\textwidth]{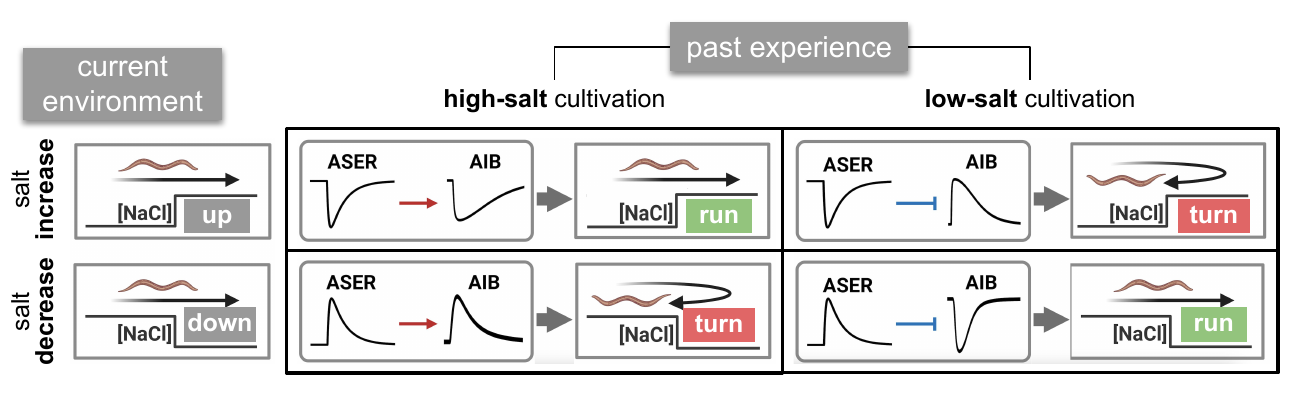}
}
\caption{\textbf{Dual AIB response to ASER enables experience-dependent behavior}.
This 2$\times$ cartoon matrix shows schematically the responses of the two neurons considered in our model (the sensory neuron ASER and the turn interneuron AIB) and the resulting behavior of the worm for two binary sets of environmental and experiential conditions (rows and columns, respectively).
Low AIB activity implies enhanced turn probability (matrix cells labeled ``turn'' in red), while
high AIB activity implies decreased turn probability (matrix cells labeled ``run'' in green).
As a result the worm always tends to move towards the salt levels that it had previously experienced.}
\label{fig:2}  
\end{figure}

The key question is how ASER drives this bidirectional reorientation behavior, i.e. how it leads to high- and low-salt attraction in high- and low-salt cultivated animals, respectively.
It has been suggested that ASER generates experience-dependent chemotaxis by altering the magnitude of excitatory and inhibitory signals to its downstream interneurons AIA, AIB, AIY, and AIZ \cite{kunitomo2013concentration, luo2014dynamic}. 
In our model, we consider for simplicity a single ASER target, namely AIB, which is by far the most studied in this context.

AIB is a turn neuron, i.e. its activation leads to an increase in the probability that the worm re-orients its crawling direction \cite{wakabayashi2004neurons, garrity2010running}. 
In that sense, here we focus on the \textit{C. elegans} locomotion strategy known as klinokinesis, consisting of a biased random walk characterized by periods of forward motion interrupted by sharp turns, in which the worm changes its orientation in a random manner \cite{iino2009parallel,izquierdo2016whole}.
Within that context, increased activity of the AIB interneuron leads to more turns, and thus to a decreasing ability to maintain the current course.
Oppositely, AIB inhibition would compel the worm to keep moving in the same direction.

How do the activities of the ASER and AIB neurons regulate the experience-dependent chemotaxis shown in Fig.~\ref{fig:1}?
A large body of experimental knowledge has been amassed in recent years \cite{kunitomo2013concentration, luo2014dynamic, ohno2017dynamics, sato2021glutamate, hiroki2022molecular} to put together a picture of what happens.
ASER is an OFF sensory neuron, i.e. it is activated (it depolarizes) upon decreases in NaCl concentration \cite{suzuki2008functional}.
This can be seen in the bottom row of the 2$\times$2 cartoon matrix shown in Fig.~\ref{fig:2}, where the activity of ASER is shown schematically in response to a down-step in salt.
ASER activation in response to such stimulus occurs in both high- and low-salt cultivation conditions (left and right columns of the cartoon matrix, respectively).
Conversely, ASER hyperpolarizes (it is inhibited) by an increase in NaCl, again under both cultivation conditions (top row in Fig.~\ref{fig:2}).
We note that the changes in ASER activity in response to salt steps need to be transient, to prevent the worm from being stuck in the same direction of motion for long periods of time.
Experiments show that the timescale of these activity changes is of the order of seconds \cite{suzuki2008functional, dekkers2021plasticity, kunitomo2013concentration}.

It is worth emphasizing from the discussion above that the response of ASER to a change in salt concentration is the same irrespective of whether the worm had been previously cultivated in high or low salt conditions (compare the behavior of ASER across a given row in the left and right columns in the matrix of Fig.~\ref{fig:2}, respectively).
How are then past experiences encoded in the system?
What changes between the two cultivation conditions is the effect that ASER has on its target neuron AIB \cite{kunitomo2013concentration, sato2021glutamate}:
under high-salt cultivation ASER \textit{activates} AIB (red arrows, left column in the matrix of Fig.~\ref{fig:2}), whereas cultivation in low salt leads to ASER \textit{inhibiting} AIB (blue blunt-end arrows, right column in the figure).
The nature of the neuronal connection does not depend on whether the current environmental conditions correspond to an up-step or a down-step of salt.
Additionally, the timescale of the changes in the nature of connection from ASER to AIB (from excitatory to inhibitory) has to be long enough to maintain a ``memory'' of the past experiences (on the order of minutes \cite{ohno2017dynamics}, not shown in the figure).

In summary, the combination of two orthogonal effects (current salt changes affecting the activity level of ASER, and past experiences affecting the way in which ASER acts upon AIB) enables combinatorially the four different actions that underlie experience-dependence salt chemotaxis in \textit{C. elegans}:
\begin{itemize}
    \item A salt up-step after high-salt cultivation leads to ASER inhibition and an excitatory connection between ASER and AIB, which in turn leads to AIB inhibition and motion forward (top left panel in the matrix of Fig.~\ref{fig:2}).
    \item A salt up-step after low-salt cultivation leads to ASER inhibition and an inhibitory connection between ASER and AIB, which in turn leads to AIB activation and increased turns (top right panel in the figure).
    \item A salt down-step after high-salt cultivation leads to ASER activation and an excitatory connection between ASER and AIB, which in turn leads to AIB activation and increased turns (bottom left panel in the figure).
    \item A salt down-step after low-salt cultivation leads to ASER activation and an inhibitory connection between ASER and AIB, which in turn leads to AIB inhibition and motion forward (bottom right panel in the figure).
\end{itemize}
These four situations lead to the expectation that the worm should be able to move towards or away from high salt concentrations depending on whether it had previously experienced high- or low-salt conditions, respectively.
However, it is still unclear whether this behavior would take place in a gradient, where the perceived changes in salt concentrations are continuous instead of consisting of finite steps.
Filling this gap can be accomplished by modeling mathematically the neural dynamics described above in an \textit{in silico} gradient, and quantifying statistically the trajectories of simulated worms undergoing biased random walks.
In order to implement this model in a biologically realistically manner, however, we first need to establish the molecular mechanisms underlying the neural dynamics described above.

\section{Molecular determinants of experience-dependent chemotaxis}

As discussed above, the plasticity of salt chemotaxis in \textit{C. elegans} relies on two types of encoding, which need to be explained molecularly: the encoding of the current environment in the activity of ASER, and the encoding of past experiences in the nature of the connection between ASER and AIB.
In what follows we discuss the two processes separately.

\subsection{Molecular encoding of current environmental conditions}
\label{sec:ffl}

We explained in Sec.~\ref{sec:neural} above that ASER responds to steps in NaCl by transiently depolarizing (hyperpolarizing) in response to a down-step (up-step) in salt.
No explicit molecular mechanisms underlying such a response have been reported yet, to our knowledge.
The ASER response to a salt step is an instance of perfect adaptation, since after the transient, the activity of the neuron goes back to the level it had prior to the change in salt concentration (even though the NaCl level is now different).
This property can be used as a constraint when searching for a potential molecular circuitry that exhibits this response \cite{ferrell2016,alon2018,khammash2021}.
In what follows we hypothesize a specific molecular mechanism based on a systematic search of the \textit{C. elegans} literature on chemosensory signal transduction, constrained by the need of explaining perfect adaptation. 

\begin{figure}[htbp]
\centerline{%
\includegraphics[width=0.5\textwidth]{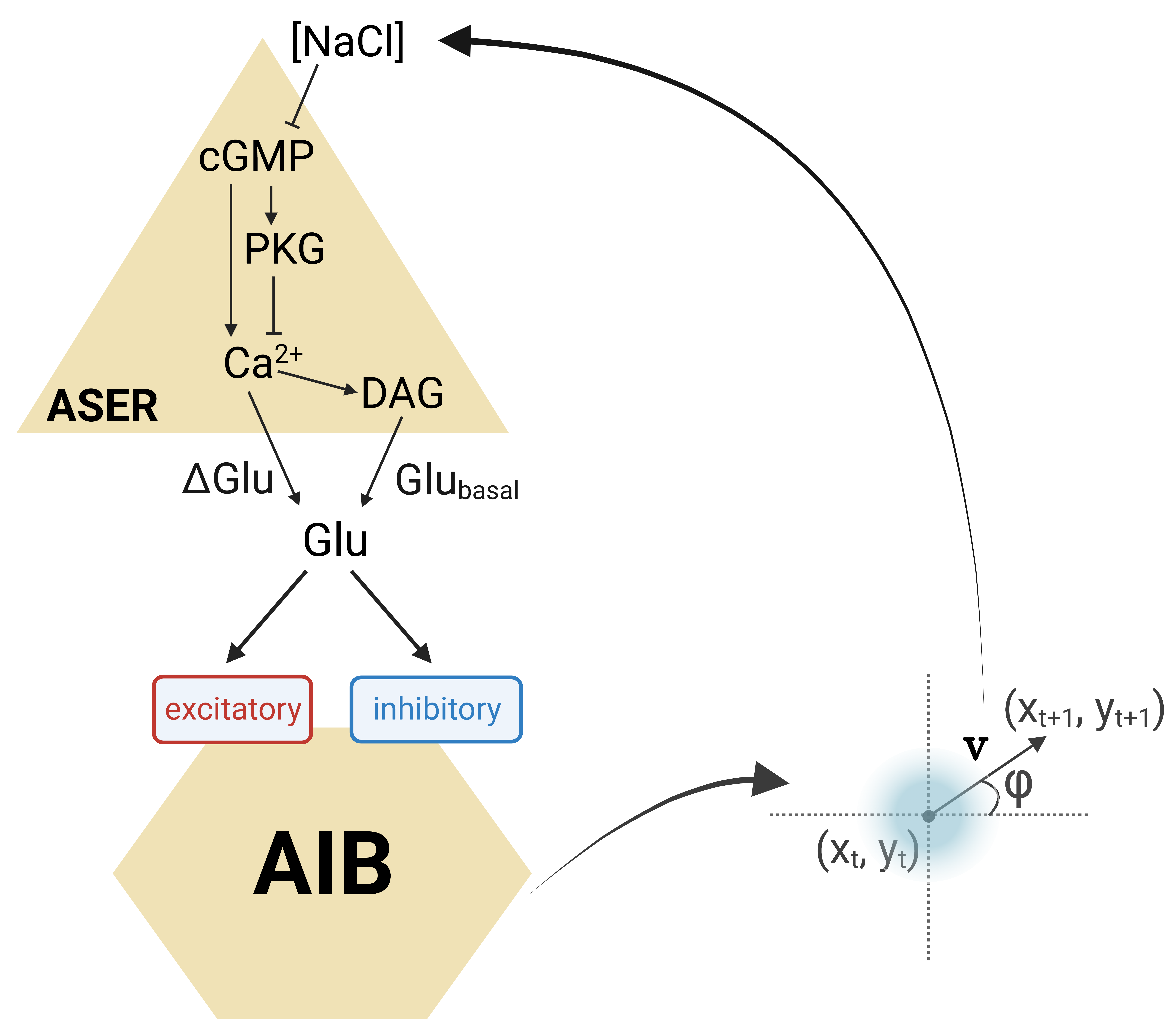}
}
\caption{\textbf{NaCl chemotaxis sensorimotor circuit}. Salt inhibits the activity of the sensory neuron ASER (triangle), which in turn affects the activity of the interneuron AIB (hexagon).
The state of AIB controls the motion of the worm by modulating the probability at which random reorientations of its motion occur.
Finally, the motor activity of the organism feeds back on the level of salt that it senses, since the salt concentration will depend on the worm's location.
The molecular components considered in our integrated model, and the interactions between them, are shown.
}
\label{fig:circuit}  
\end{figure}

ASER senses salt stimuli through guanylyl cyclase receptors (rGCs) \cite{ferkey2021chemosensory}. 
These membrane proteins consist of two parts: an extracellular receptor domain and an intracellular guanylyl cyclase activity domain.
When salt binds the extracellular domain of rGC, cyclase activity is inhibited and signaling is silenced.
Conversely, signaling is initiated when salt is removed and the intracellular cyclase domain induces the production of cyclic guanosine monophosphate (cGMP) \cite{ferkey2021chemosensory}.
cGMP binds to and opens cGMP-gated calcium channels in ASER, leading to an influx of calcium ions, and consequently to neuronal depolarization. 
In parallel, cGMP activates PKG (a cGMP-dependent protein kinase, also called EGL-4 in the \textit{C. elegans} literature).
PKG is known to be required in ASER for calcium signals to arise in response to salt \cite{ferkey2021chemosensory}, but its targets and role are unknown.
One of the proposed hypothesis is that PKG inhibits cGMP-gated channels, thereby inhibiting the influx of calcium ions \cite{ferkey2021chemosensory}.
Together, these interactions take the form of an incoherent feedback circuit between cGMP, PKG and Ca$^{2+}$, with NaCl acting as upstream input.
This is shown in Fig.~\ref{fig:circuit} (molecular circuit inside the triangle representing ASER, top of the figure).
Incoherent feedforward circuits are among a handful of molecular circuits that can exhibit perfect adaptation \cite{ferrell2016,alon2018,khammash2021}, which makes this hypothesis potentially attractive.
In the following section we will include this feedforward circuit in our integrated model of experience-dependent behavior.

\subsection{Molecular encoding of past experiences}
\label{sec:past}

\begin{figure}[htbp]
\centerline{\includegraphics[width=0.3\textwidth]{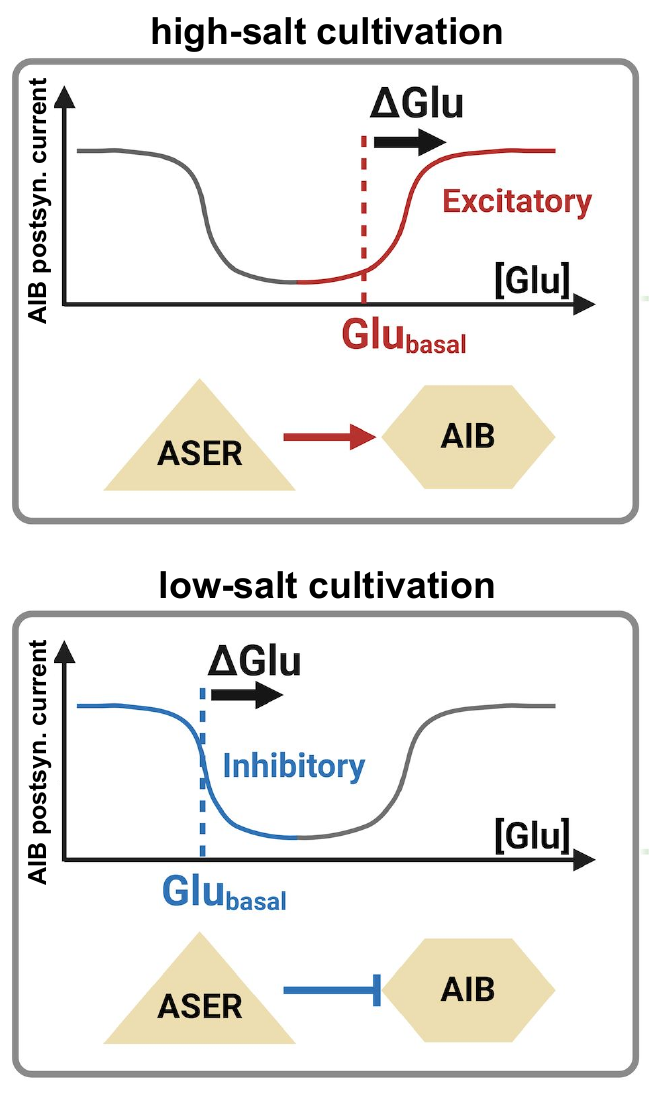}}
\caption{\textbf{Dual regulation of AIB by ASER encodes past experiences.}
Response of AIB in terms of its postsynaptic current as a function of glutamate levels.
Top: under high-salt cultivation conditions, the synapse operates in the right-hand side of the U curve, thus acting in an excitatory manner.
Bottom: under low-salt cultivation conditions, the synapse operates in the left-hand side of the U curve, thus acting in an inhibitory manner.
}
\label{fig:AIB}  
\end{figure}

The experience-dependent change of synapse polarity between ASER and AIB shown schematically in Fig.~\ref{fig:2} is known to arise through glutamate signaling.
ASER releases glutamate upon activation, which is sensed by AIB through two types of glutamate receptors: an excitatory glutamate-gated cation channel and an inhibitory glutamate-gated chloride channel \cite{sato2021glutamate, hiroki2022molecular}.
Crucially, these two receptors have different sensitivities: the inhibitory receptor has a smaller glutamate threshold than the excitatory one \cite{hiroki2022molecular}.
This leads to a U-shaped response of AIB to glutamate, as shown in Fig.~\ref{fig:AIB}.

Given this non-monotonic response of AIB to ASER, the worm can set the nature of the synaptic connection to be either excitatory or inhibitory by operating in high or low glutamate concentration ranges, respectively.
Recently it was proposed, on the basis of experimental observations, that this can be achieved by controlling the level of basal glutamate release by ASER \cite{hiroki2022molecular}.
This strategy is based on the way in which synaptic transmission occurs in \textit{C. elegans}: unstimulated sensory neurons release a basal amount of neurotransmitter, which gradually increases or decreases based on sensory perception \cite{ventimiglia2017diverse, narayan2011transfer, lindsay2011optogenetic}. 

In our case, when basal glutamate is low, the synapse operates in a predominantly inhibitory mode (blue section of the response curve in Fig.~\ref{fig:AIB}, bottom panel). Conversely, when basal glutamate is high, the effect of ASER's additional glutamate release will be excitatory (red section of the postsynaptic current curve in Fig.~\ref{fig:AIB}, top panel). 
The intriguing question now is how the level of basal glutamate released by ASER is regulated in an experience-dependent manner: for high-salt cultivated animals the basal release should be high, while under low-salt cultivation basal glutamate should be low.

Glutamate release by \textit{C. elegans} sensory neurons is known to be up-regulated by the DAG/PKC-1 signalling pathway \cite{ventimiglia2017diverse}.
In particular, DAG/PKC-1 has been found to up-regulate glutamatergic transmission by ASER \cite{hiroki2022molecular}. 
Furthermore, it has been shown that DAG levels in ASER are dynamically regulated in response to step changes in NaCl concentration \cite{ohno2017dynamics}, with DAG increasing (decreasing) transiently in a response to down-step (up-step) in salt.
Taken together, these features suggest a possible explanation of how basal glutamate release is regulated in an experience-dependent manner.
Specifically, high-salt cultivated animals face a down-step in salt, and thus a transient increase in DAG, when placed in the middle of a gradient \cite{ohno2017dynamics}.
Such an increase in DAG would produce an increase in basal glutamate, leading to an excitatory connection between ASER and DAG, as expected (left column in Fig.~\ref{fig:2} and top panel in Fig.~\ref{fig:AIB}).
Conversely, for low-salt cultivated animals NaCl concentration is increased when transferred to the gradient plate, which lowers DAG levels \cite{ohno2017dynamics}.
This decreases basal glutamate release, leading to an inhibitory postsynaptic response in AIB (right column in Fig.~\ref{fig:2} and bottom panel in Fig.~\ref{fig:AIB}).
Notably, DAG levels are maintained for tens of minutes in the ASER neuron \cite{ohno2017dynamics}, which coincides with the time span necessary to guide the worm towards the cultivation concentration \cite{hiroki2022molecular}. 

The molecular and cellular processes described above seem to explain experience-dependent salt chemotaxis in \textit{C. elegans}.
On the other hand, the connection between the two modes of encoding is still unclear, as is the extrapolation of responses to step changes in salt to a gradient, as mentioned in Sec.~\ref{sec:neural} above.
To put all the ingredients together and bridge the gap between molecular/cellular mechanisms and behavior, we now implement an integrated mathematical model that bridges the molecular, cellular and organism scales.

\section{An integrated model of experience-dependence behavior}

In this section we describe how we built our mathematical model based on the previously described experimental observations.
As depicted in Figure \ref{fig:circuit}, our model includes a sensory neuron, ASER, that affects the activity of an interneuron, AIB, which in turn regulates the motor system of the worm.
This cellular description is complemented at the molecular level by specific circuits regulating the responses of the two neurons, and at the behavioral level by a description of how the activity of AIB affects the motor output of the organism, which leads to motion changes (or lack thereof) in a gradient, and thus to changes in the sensed NaCl, closing the loop.

\subsection{An intracellular sensory feedforward circuit}
\label{sec:ffl-model}

We begin with the dynamics of the sensory neuron ASER.
As mentioned above, ASER produces Ca$^{2+}$ transients in response to NaCl concentration changes \cite{suzuki2008functional}. Our aim is to model this response with a biologically grounded yet simple circuit.
We have not found in the literature this kind of model for ASER, or for any other \textit{C. elegans} sensory neuron.
There are effective models that use explicit derivative-like operations to account for the transduction of a salt step into a Ca$^{2+}$ pulse \cite{izquierdo2010evolution, izquierdo2013connecting}.
The advantage of those frameworks is that they are easy to construct and tune, but they do not provide a molecular basis for the derivative-like operations mentioned above. 
On the other hand, there are physiologically grounded models for \textit{C. elegans} sensory neurons, such as the neurons AWC \cite{usuyama2012model} and ASH \cite{mirzakhalili2018mathematical}, but they are overly detailed models, which are difficult to tune and lack experimental evidence to support some of the assumptions made and the parameter values. 

\begin{figure}[htbp]
\centering
(a)
\begin{minipage}[t]{0.40\textwidth}
\vspace{0pt}
\includegraphics[width=\textwidth]{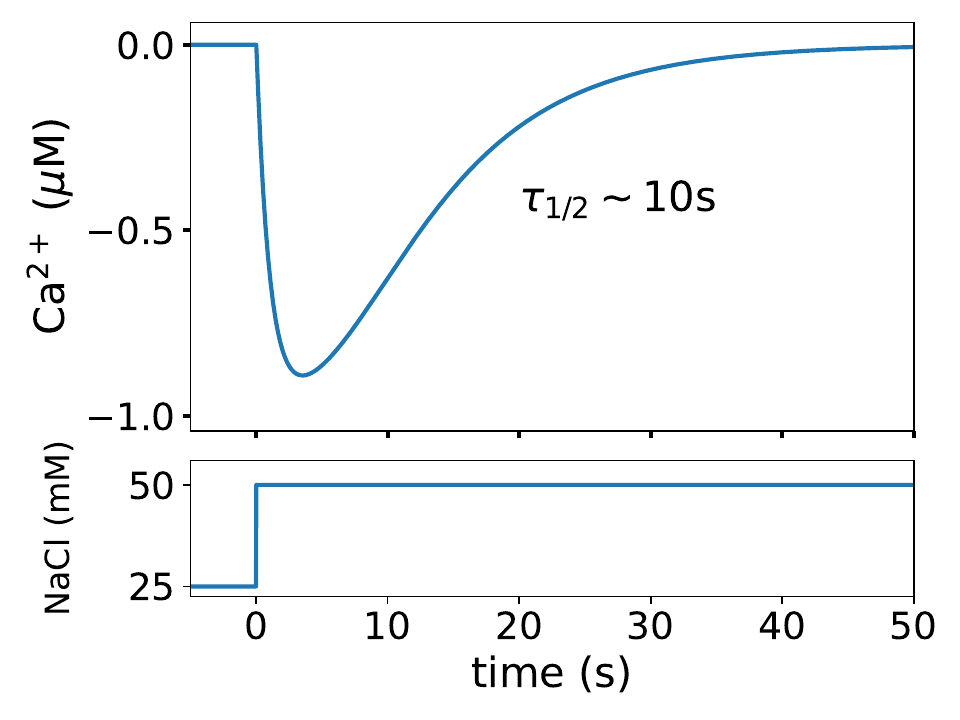}
\end{minipage}
(b)
\begin{minipage}[t]{0.40\textwidth}
\vspace{0pt}
\includegraphics[width=\textwidth]{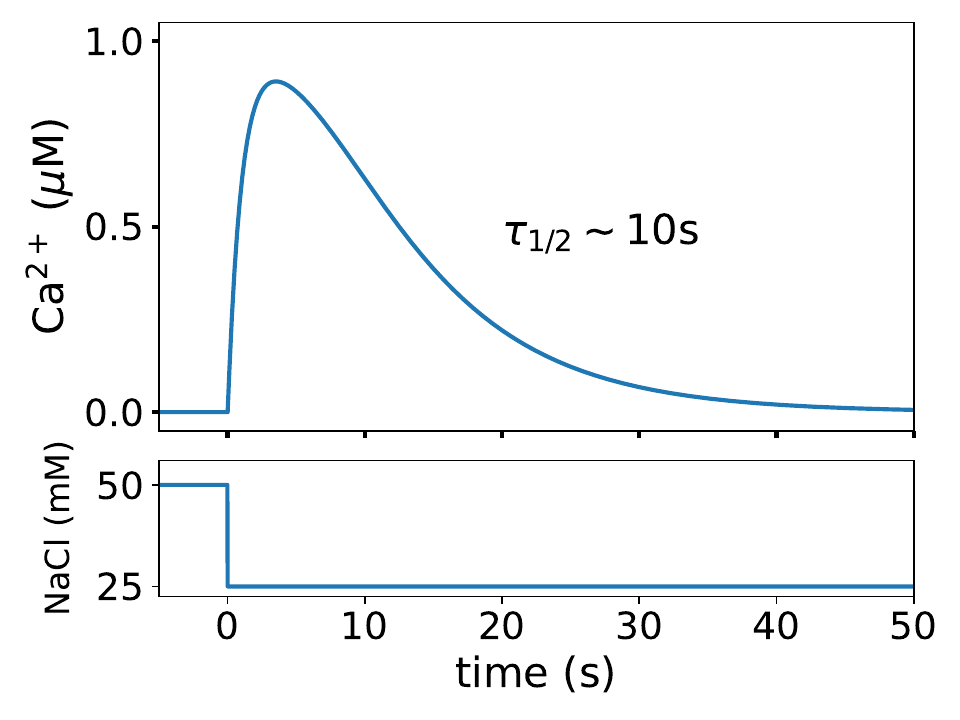}
\end{minipage}
\label{upstep_ca}
\caption{\textbf{Simulated response of ASER's activity to NaCl steps}.
The intracellular $\mathrm{Ca}^{2+}$ concentration of ASER is shown for a NaCl upstep (a) and downstep (b).
The response time of the model is in agreement with the one measured experimentally \cite{hiroki2022molecular} ($\tau_{\mathrm{1/2}}\sim 10$~s).}
\label{ca_steps}  
\end{figure}

In that context, a parsimonious framework is missing, biologically-grounded and yet not too complex.
The feedforward circuit proposed in Sec.~\ref{sec:ffl} above is an adequate choice for such a model.
As we described there (see also Fig.~\ref{fig:circuit}), cGMP activation in ASER affects Ca$^{2+}$ influx in two ways: directly via the opening of cGMP-gated channels (leading to an increase in Ca$^{2+}$ influx), and indirectly via the activation of the cGMP-gated channel inhibitor PKG (leading to a decrease in Ca$^{2+}$ influx). 
The dynamics of this incoherent feedforward circuit can be represented by the following system of coupled differential equations:
\begin{align}
 &\frac{d \mathrm{cGMP}}{d t} = \frac{\alpha}{1+\mathrm{[NaCl]}/K} -\delta_{\mathrm{GMP}}\cdot\mathrm{cGMP},
\label{cgmp}
\\
&\frac{d \mathrm{PKG}}{d t} = \gamma\cdot\mathrm{cGMP} - \delta_{\mathrm{PKG}}\cdot\mathrm{PKG},
\label{pkg}
\\
&\frac{d \mathrm{Ca}^{2+}}{d t} = \beta\cdot\sigma (\mathrm{cGMP} - \mathrm{PKG}) -\delta_{\mathrm{Ca}}\cdot\mathrm{Ca}^{2+}
\label{calcium}
\end{align}
The first term in Eq.~(\ref{cgmp}) represents the inhibition of rGC-mediated cGMP production by NaCl, as discussed in Sec.~\ref{sec:ffl} above \cite{ferkey2021chemosensory}.
To model this we have chosen a standard repressing Michaelis-Menten function \cite{alon2006introduction}.
The first term in Eq.~(\ref{pkg}) represents the activation of PKG by cGMP, assumed to be linear.
Equation~\eqref{calcium} corresponds to the dynamics of the variable $\mathrm{Ca}^{2+}$, which represents the change in calcium levels relative to its baseline (thus Ca$^{2+}$ can assume positive and negative values).
In particular, the first term in Eq.~\eqref{calcium} represents the calcium influx due to the opening of cGMP-gated channels, which are inhibited by PKG.
According to this expression, calcium influx increases (with respect to its baseline) when cGMP surpasses PKG according to a threshold function given by the sigmoid function
\begin{equation}
 \sigma(x) = \tanh (b x)\,.
\label{tanh}
\end{equation}
The last terms in Eqs.~(\ref{cgmp})-(\ref{calcium}) represent the decay of the three species.
The values of these decay rates (see Table~\ref{tab:param}) are chosen such that the timescale of calcium dynamics is fast enough (on the order of seconds) to avoid locking the organism in a fixed direction of motion for too long.
The resulting time traces of Ca$^{2+}$ levels inside ASER in response to NaCl steps are shown in Fig.~\ref{ca_steps}.
All parameter values can be found in Table~\ref{tab:param}. 

Figure \ref{ca_steps} shows that ASER depolarizes (its $\mathrm{Ca}^{2+}$ levels increase) in response to a NaCl decrease, and hyperpolarizes when NaCl increases.
Experimentally, the $\mathrm{Ca}^{2+}$ response to a NaCl downstep is more pronounced than to an upstep \cite{suzuki2008functional}.
However, in our model the responses are considered symmetric, as show in Fig. \ref{ca_steps}.
This simplification has already been used in other models of \textit{C. elegans} navigation \cite{dekkers2021plasticity}, for the same parsimonious reasons as ours.
The parameters of Eqs.~\eqref{cgmp}-\eqref{tanh} were chosen to match the experimental response time of ASER's $\mathrm{Ca}^{2+}$ levels upon a NaCl decrease (defined as the time to reach half the distance between the peak and resting level) \cite{hiroki2022molecular}.
Specifically, the response time to a 25~mM NaCl step is $\sim 10$s, as depicted in Fig.~\ref{ca_steps}.

\subsection{A dual neuronal signaling process}
\label{sec:dual}

\begin{figure}[htbp]
\centering
(a)
\begin{minipage}[t]{0.40\textwidth}
\vspace{0pt}
\includegraphics[width=\textwidth]{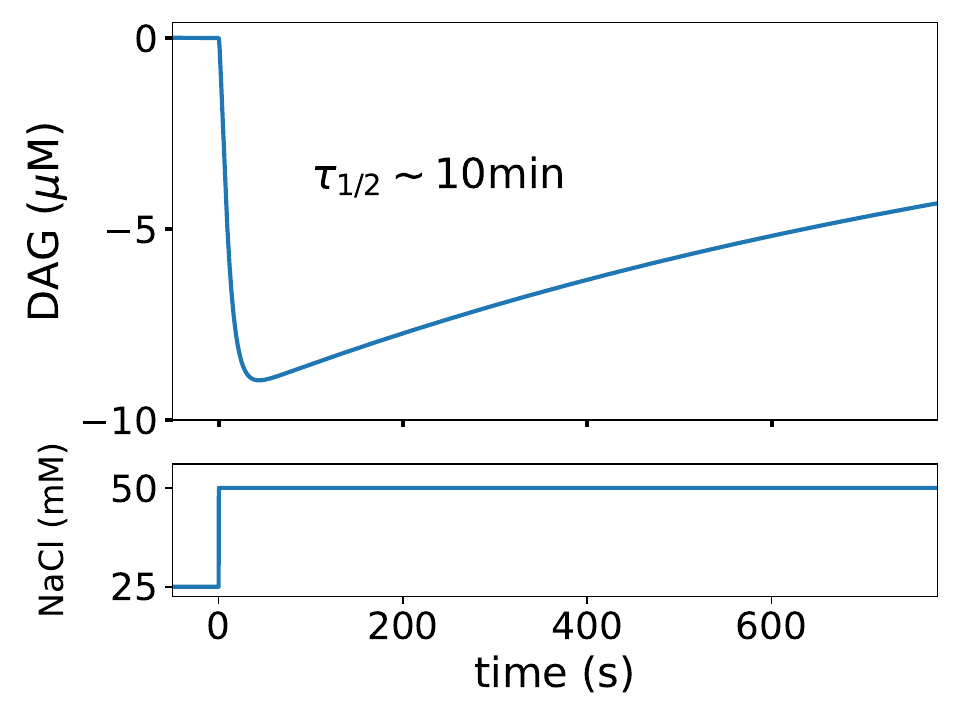}
\end{minipage}
(b)
\begin{minipage}[t]{0.40\textwidth}
\vspace{0pt}
\includegraphics[width=\textwidth]{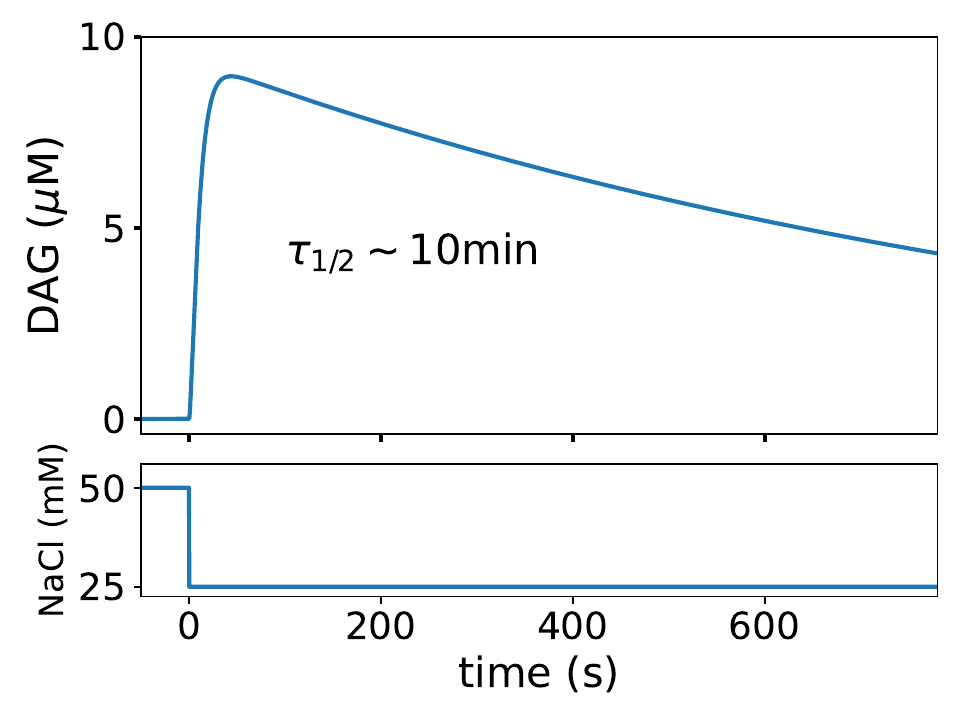}
\end{minipage}
\caption{\textbf{Simulated response of DAG levels to NaCl steps.}
The intracellular DAG concentration in ASER is shown for a NaCl upstep (a) and downstep (b).
The model response time is in agreement with the measured in Ref. \cite{ohno2017dynamics} ($\tau_{\mathrm{1/2}}\sim 600$~s).}
\label{DAG_steps}  
\end{figure}

The response of DAG to changes in NaCl concentration mentioned in Sec.~\ref{sec:past} above has been shown to be caused to changes in the neural activity of ASER \cite{ohno2017dynamics}.
Specifically, ASER's intracellular $\mathrm{Ca}^{2+}$ enhances DAG production, most likely via the PLC-$\beta$/EGL-8 pathway \cite{hiroki2022molecular}.
We represent this effect in our model via the following equation: 
\begin{equation}
 \frac{d\mathrm{DAG}}{dt} = \beta_{\mathrm{DAG}}\cdot\mathrm{Ca}^{2+} -\delta_{\mathrm{DAG}}\cdot\mathrm{DAG},
\label{dag_dyn}
\end{equation}
where DAG up-regulation by $\mathrm{Ca}^{2+}$ is modeled as a linear activation, and the variable represents changes in DAG levels relative to its baseline (and thus it can assume positive or negative values).
The simulated responses of DAG activity in the ASER neuron for both a salt upstep and downstep are shown in Fig.~\ref{DAG_steps}.

Figure~\ref{DAG_steps} shows that DAG levels decrease upon a sudden NaCl increase (panel a), while they rise upon a sudden NaCl decrease (panel b), in both cases mediated by $\mathrm{Ca}^{2+}$.
Similarly to the case of the sensory feedforward circuit discussed in Sec.~\ref{sec:ffl-model} above, the parameters of Eq.~\eqref{dag_dyn} were chosen to match the experimentally observed decay time of DAG \cite{ohno2017dynamics, hiroki2022molecular}, which is $\sim$10~min. 

The next step in our model is to describe the effect of the sensory neuron ASER on the turn neuron AIB.
As discussed in Sec.~\ref{sec:past}, ASER communicates with AIB synaptically via the neurotrasmitter glutamate.
The glutamate released by ASER has two components: a slow basal level $\mathrm{Glu_{basal}}$ controlled by DAG, and a fast instantaneous level $\Delta \mathrm{Glu}$ controlled by the intracellular calcium concentration of ASER \cite{hiroki2022molecular}:
 \begin{align}
  \mathrm{Glu} &= \beta_{\mathrm{Glu}} + \mathrm{Glu_{basal}} + \Delta \mathrm{Glu}\\ 
   &= \beta_{\mathrm{Glu}} + \alpha_{\mathrm{Glu}} \mathrm{H}(\mathrm{DAG} - \theta) + \alpha_{\Delta}\mathrm{Ca}^{2+},
\label{glutamate}
 \end{align}
where $\beta_{\mathrm{Glu}}$ is a (small) background glutamate level.
We assume that basal glutamate is produced when DAG exceeds a threshold value $\theta$, as described by the Heaviside step function $\mathrm{H(x)}$ (which is 1 when $x\ge 0$ and 0 otherwise).
The instantaneous excess glutamate, $\Delta \mathrm{Glu}$, is assumed in turn to increase linearly with the intracellular $\mathrm{Ca}^{2+}$ levels of ASER. It is worth mentioning that $\mathrm{Glu_{basal}}$ and $\Delta \mathrm{Glu}$ have different timescales, corresponding to those of their activating factors: $\mathrm{Glu_{basal}}$, regulated by DAG, has a characteristic time $\sim 10$ minutes \cite{ohno2017dynamics}, whereas $\Delta \mathrm{Glu}$ operates in time scales on the order of $\sim 10$ seconds, corresponding to the $\mathrm{Ca}^{2+}$ dynamics of ASER \cite{suzuki2008functional, dekkers2021plasticity, kunitomo2013concentration}. 

\begin{figure}[htbp]
\centering
(a)
\begin{minipage}[t]{0.40\textwidth}
\vspace{0pt}
\includegraphics[width=\textwidth]{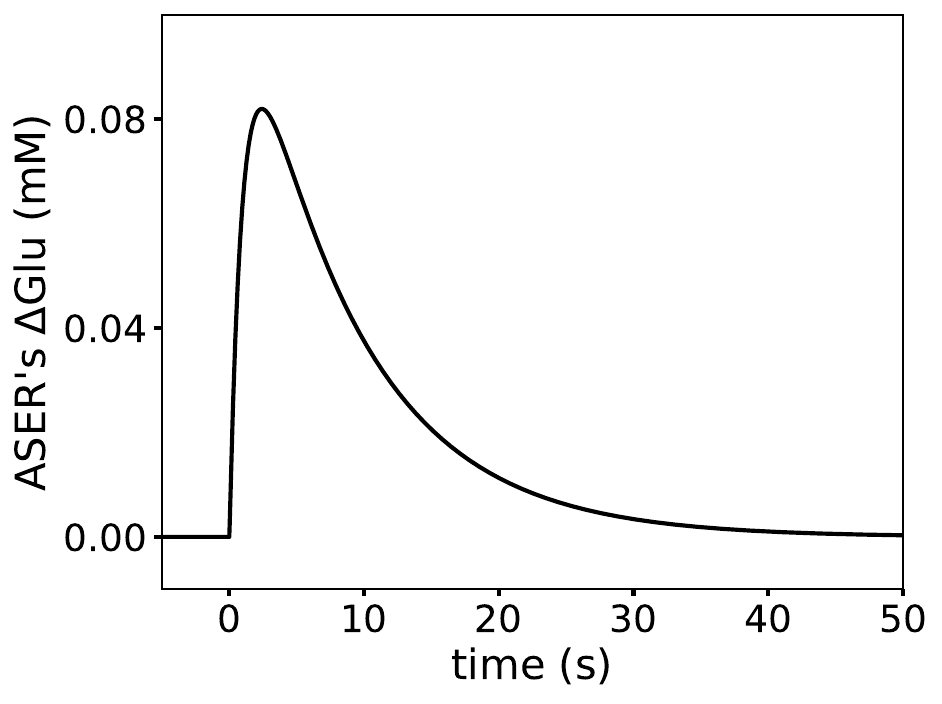}
\end{minipage}
(b)
\begin{minipage}[t]{0.40\textwidth}
\vspace{0pt}
\includegraphics[width=\textwidth]{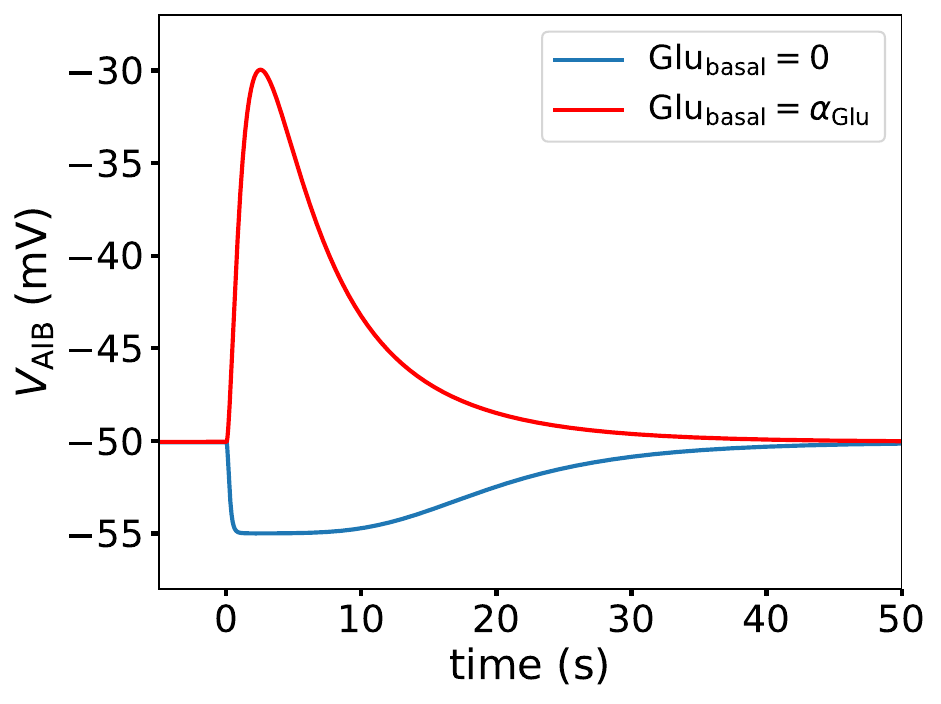}
\end{minipage}
\caption{\textbf{Experience-dependent $V_{\mathrm{AIB}}$ response.}
(a) Instantaneous change of glutamate released by ASER ($\Delta \mathrm{Glu}=\alpha_{\Delta}\mathrm{Ca}^{2+}$) in response to a NaCl downstep.
(b) corresponding AIB response to the $\Delta \mathrm{Glu}$ depicted in the left panel.
Blue curve: inhibitory response for high salt cultivation ($\mathrm{Glu_{basal}}=0$).
Red curve: excitatory response for low salt cultivation ($\mathrm{Glu_{basal}} > 0$).}
\label{pre_post} 
\end{figure}

The total glutamate released by ASER acts upon AIB via the excitatory and inhibitory synapses discussed in Sec.~\ref{sec:neural}, such that the dynamics of AIB's membrane potential is given by:
 \begin{align}
 \tau \frac{d V_{\mathrm{AIB}}}{d t} = \omega_{\mathrm{inh}} S^{\mathrm{inh}} ( \mathrm{Glu}) & + \omega_{\mathrm{exc}}S^{\mathrm{exc}} (\mathrm{Glu}) \\
 &- (V_{\mathrm{AIB}}-V_{\mathrm{rest}}),
\label{vaib}
\end{align}
where the first and second terms in the right-hand side represent the inhibitory and excitatory currents, respectively \cite{sato2021glutamate}, with $\omega_{\mathrm{inh}}$ and $\omega_{\mathrm{exc}}$ denoting the weights of the corresponding connections.
The synaptic currents are assumed to depend on glutamate through the sigmoid functions
\begin{align}
    S^{\mathrm{inh}}(x) &= \frac{e^{-b_{\mathrm{inh}}(x - \theta_{\mathrm{inh}})}}{1+e^{-b_{\mathrm{inh}}(x - \theta_{\mathrm{inh}})}}
\label{s_inh}
\\
    S^{\mathrm{exc}}(x) &= \frac{1}{1+e^{-b_{\mathrm{exc}}(x - \theta_{\mathrm{exc}})}},
\label{s_exc}
\end{align}
where $b_j$ and $\theta_j$ ($j= \mathrm{exc}, \mathrm{inh}$) correspond to the steepness and threshold of each synaptic function.
The inhibition threshold is smaller than the excitation threshold (see Table~\ref{tab:param}), and as a result the overall response of $V_{\mathrm{AIB}}$ to glutamate is U-shaped, as shown in Fig.~\ref{fig:AIB} above.
This enables two different responses to the fast transient pulse in glutamate, $\Delta \mathrm{Glu}$, caused by present changes in NaCl such as a downstep (Fig.~\ref{pre_post}a). 
For low-salt cultivated animals $\mathrm{Glu_{basal}}$ is 0, so the postsynaptic response is inhibitory (Fig.~\ref{pre_post}b, blue curve). On the contrary, for high-salt cultivated animals $\mathrm{Glu_{basal}}$ is high, giving rise to an excitatory postsynaptic response (red curve).

\subsection{Effect on the worm's motion}

Finally we need to couple the molecular and cellular dynamics described about to the worm's motion.
To that end we take into account that AIB activation determines motor output, by producing instantaneous turning events that underlie the worm's pirouettes \cite{pierce1999fundamental}.
In this sense, our model assumes that during locomotion the body follows the head, allowing us to focus on the sensory-motor control of a point worm.
At each point in time the worm is modeled as a point at coordinates ($x(t)$, $y(t)$), with its head being directed towards an angle $\varphi(t)$, as depicted in Fig. \ref{fig:circuit} and expressed mathematically by
\begin{equation}
 \frac{d (x,y)}{d t} = (v\cos \varphi, v\sin \varphi),
\label{position}
\end{equation}
where we assume that the worm moves at a fixed speed $v$.

Pirouettes are executed by resetting the orientation angle $\varphi$. The probability of a pirouette per unit time $P_{\Omega}$ is determined by the activation of the AIB turn neuron.
We represent this by the following piece-wise monotonically increasing function of $V_{\mathrm{AIB}}$:
\begin{equation}
P_{\Omega } = 
  \begin{cases}
  \omega_{\mathrm{low}} &\quad \mathrm{if~}V_{\mathrm{AIB}}\leq V_{\mathrm{low}},\\
 \omega_{\mathrm{high}} &\quad \mathrm{if~}V_{\mathrm{AIB}}> V_{\mathrm{low}}, \\ 
  \end{cases}
\label{pirouette} 
\end{equation}
where the parameters $\omega_{\mathrm{low}}$ and $\omega_{\mathrm{high}}$ represent base pirouette rates. 
When a pirouette is executed, the heading $\varphi$ is instantaneously set to a random angle, drawn from a uniform distribution between $0$ and $2\pi$.

When the worm's position changes according to Eq.~\eqref{position}, \textit{C. elegans} will sense a new NaCl concentration value, which will serve as new sensory input to ASER neuron (see Eq.~\eqref{cgmp} and curved arrows in Fig.~\ref{fig:circuit}).
We now put all these pieces together to model the behavior of \textit{C. elegans} in a NaCl concentration gradient for our two different past experiences (high and low salt cultivation).

\begin{figure}[htbp]
\centering
 \includegraphics[width=0.4\textwidth]{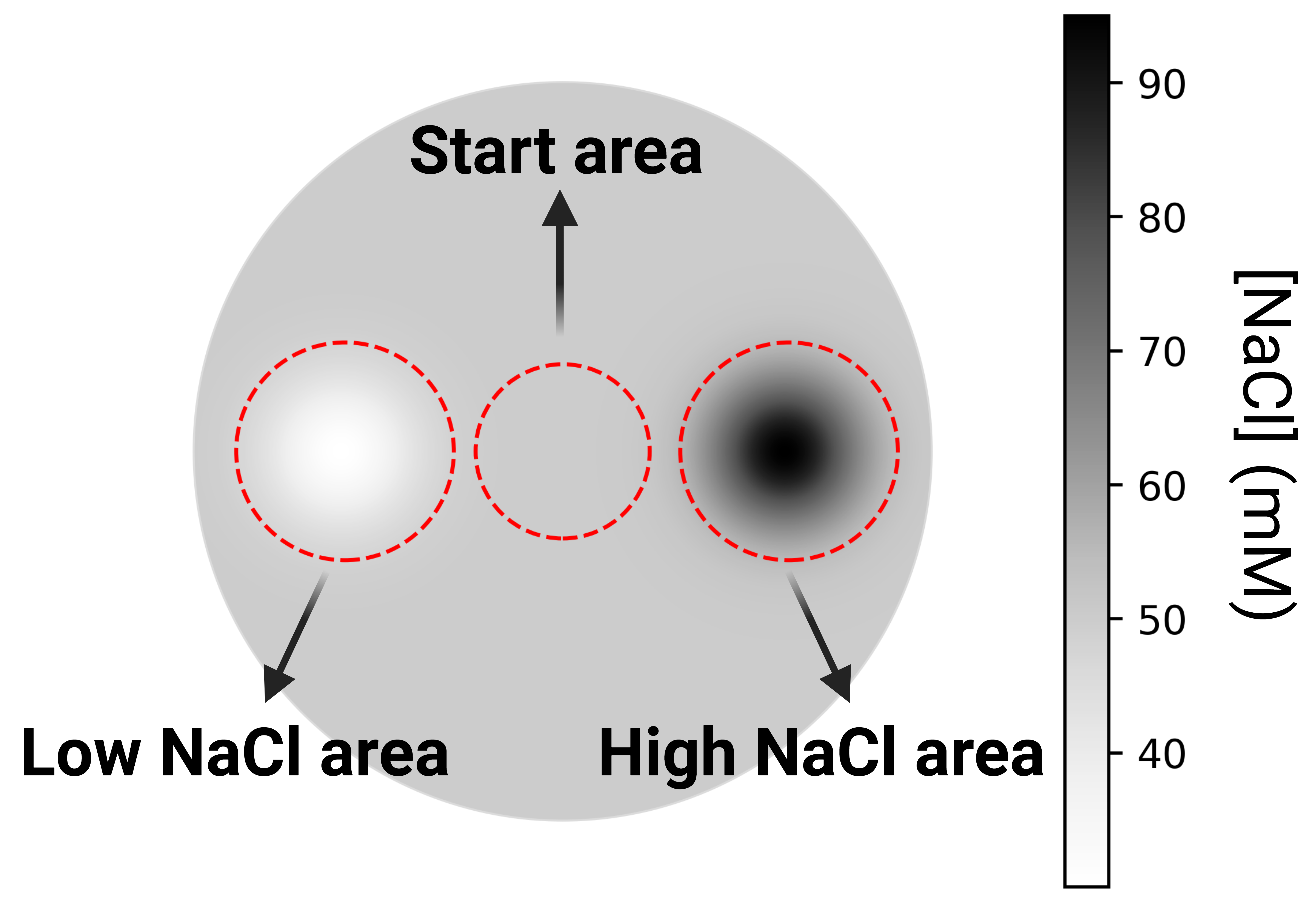}
\caption{\textbf{Chemotaxis assay.} Virtual agar plate with NaCl gradient used in our simulations.
The virtual worms are placed initially inside the start area, and the number of worms that reach the low and high NaCl areas by the end of the simulation are counted.}
\label{gradiente_plate}  
\end{figure}

\begin{figure*}[htbp]
\centering
(a)
\begin{minipage}[t]{0.28\textwidth}
\vspace{0pt}
\includegraphics[width=\textwidth]{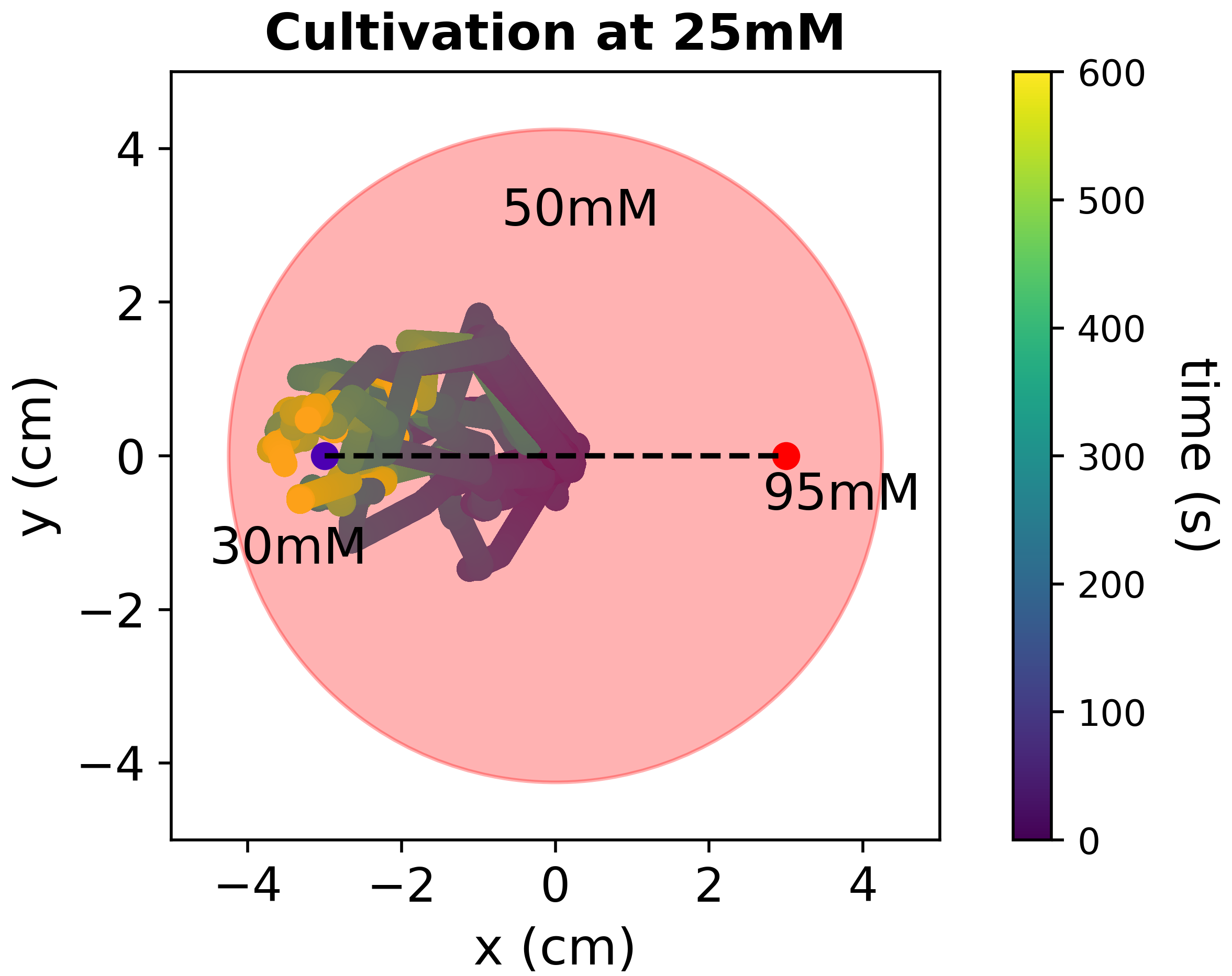}
\end{minipage}
(b)
\begin{minipage}[t]{0.28\textwidth}
\vspace{0pt}
\includegraphics[width=\textwidth]{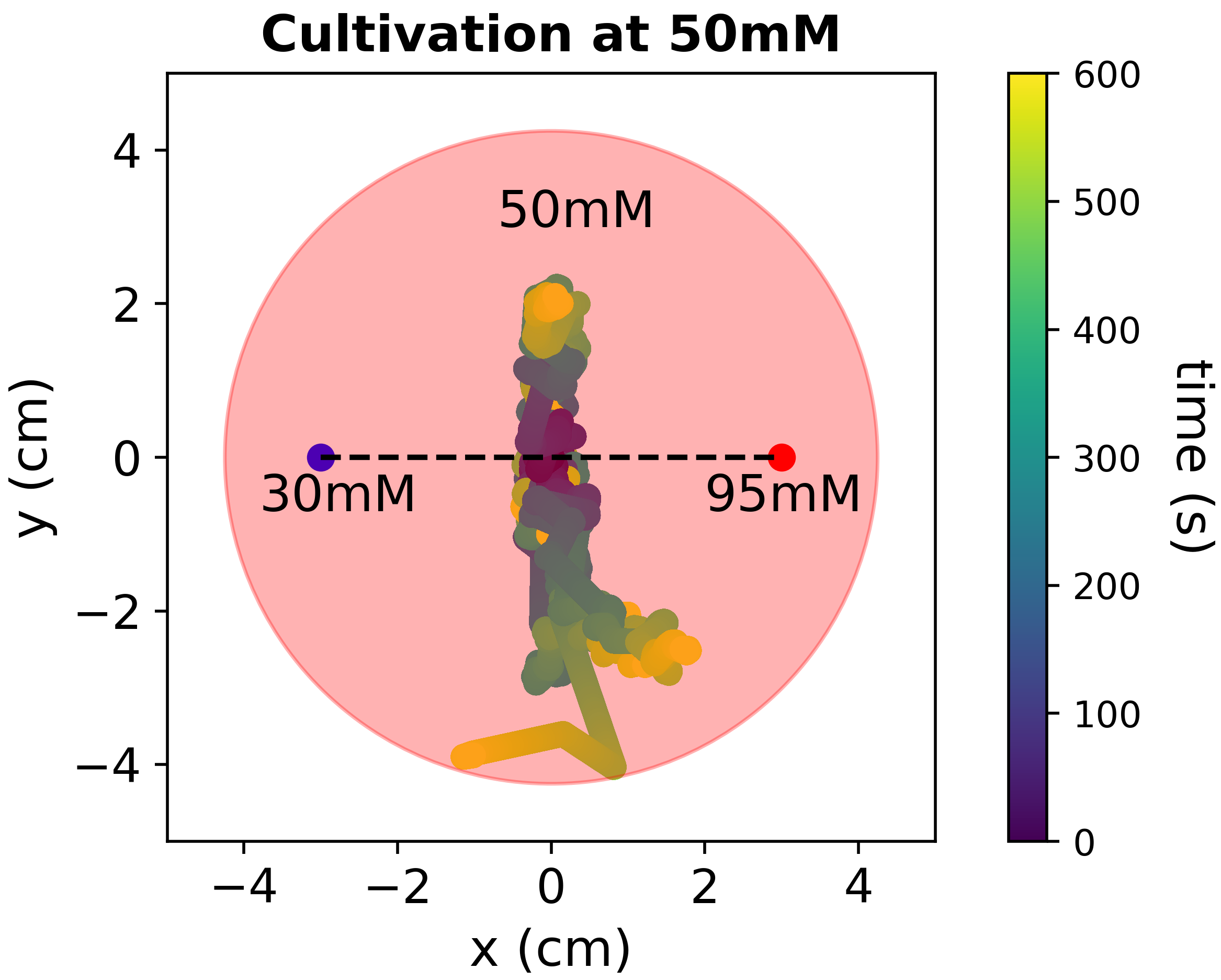}
\end{minipage}
(c)
\begin{minipage}[t]{0.28\textwidth}
\vspace{0pt}
\includegraphics[width=\textwidth]{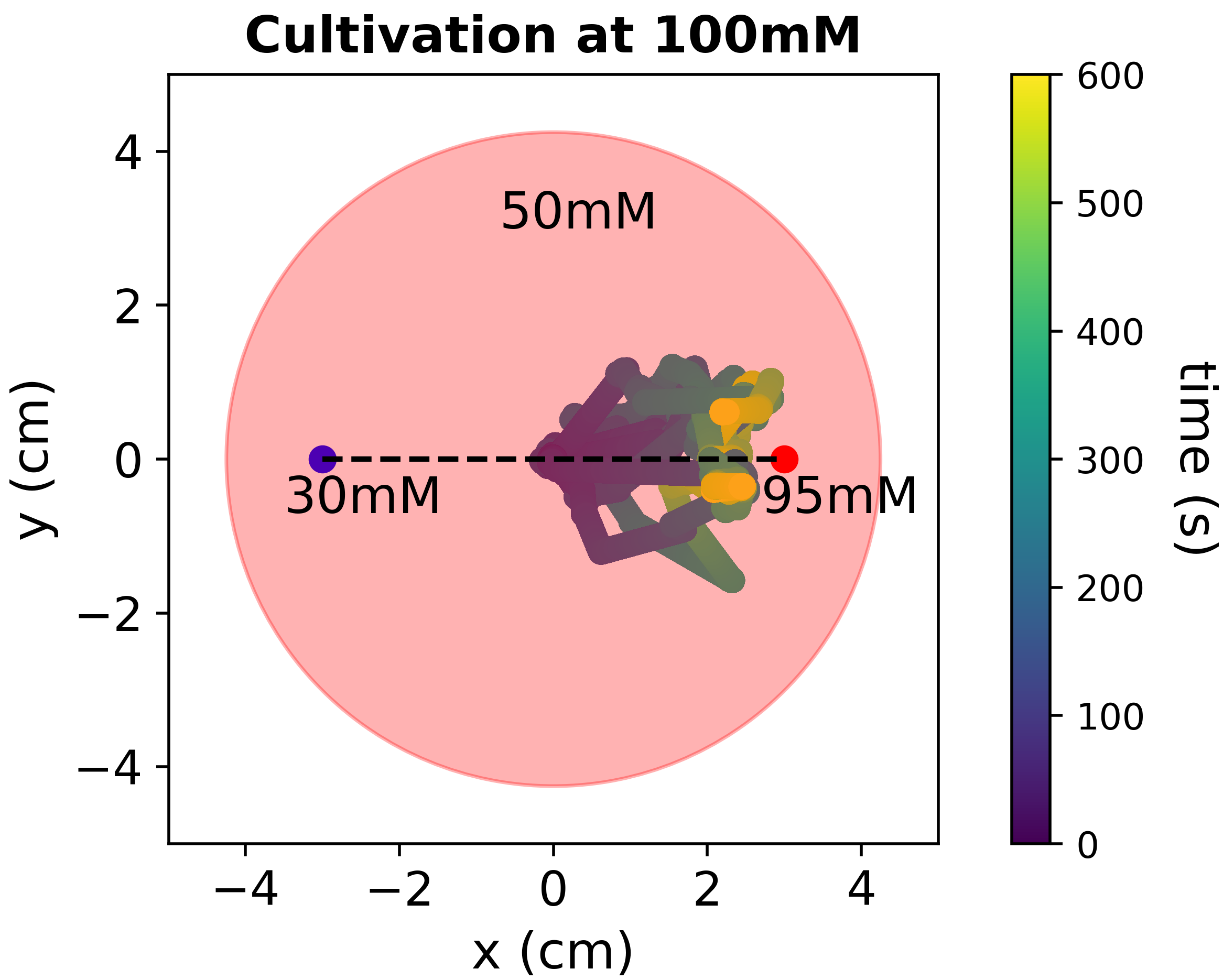}
\end{minipage}
\caption{\textbf{Behavioral traces}. Trajectories of ten virtual worms during the chemotaxis assay. Animals are placed at the center of the gradient plate and allowed to move for 600~s. Trace color represents time. (a) Worms cultivated at low [NaCl] (25~mM) move towards the low NaCl area ($\sim$30~mM). (b) Worms cultivated at the background [NaCl] of the gradient plate (50~mM) move around areas where [NaCl]=50~mM. (c) Worms cultivated at high [NaCl] (100~mM) migrate to the high NaCl area ($\sim$95~mM).}
\label{trajectories}  
\end{figure*}

\begin{figure}[htbp]
\centering
(a)
\begin{minipage}[t]{0.40\textwidth}
\vspace{0pt}
\includegraphics[width=\textwidth]{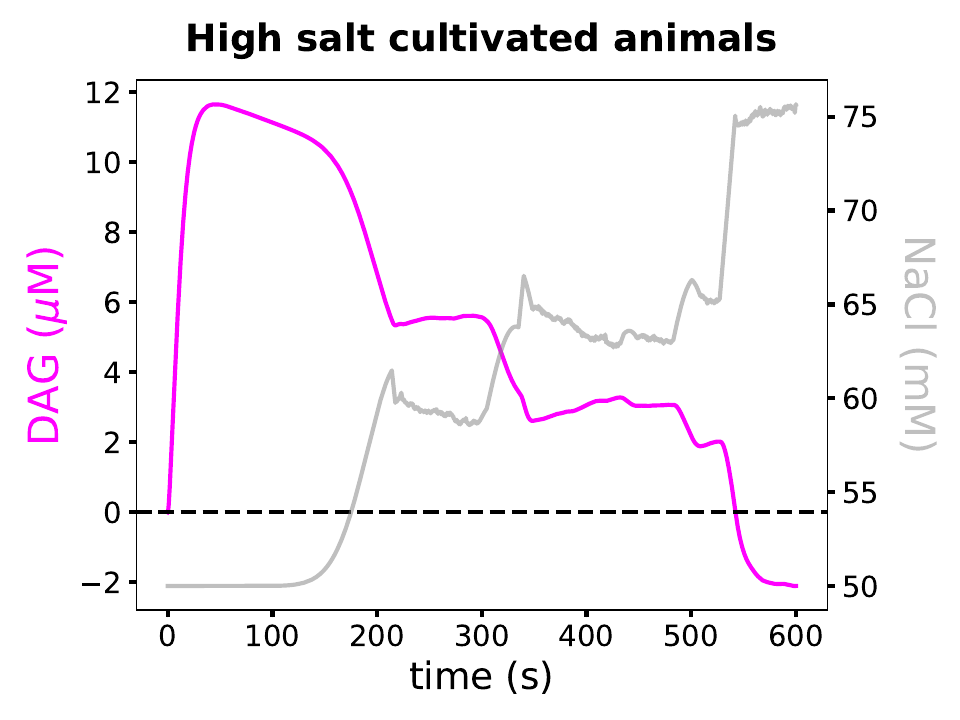}
\end{minipage}
(b)
\begin{minipage}[t]{0.40\textwidth}
\vspace{0pt}
\includegraphics[width=\textwidth]{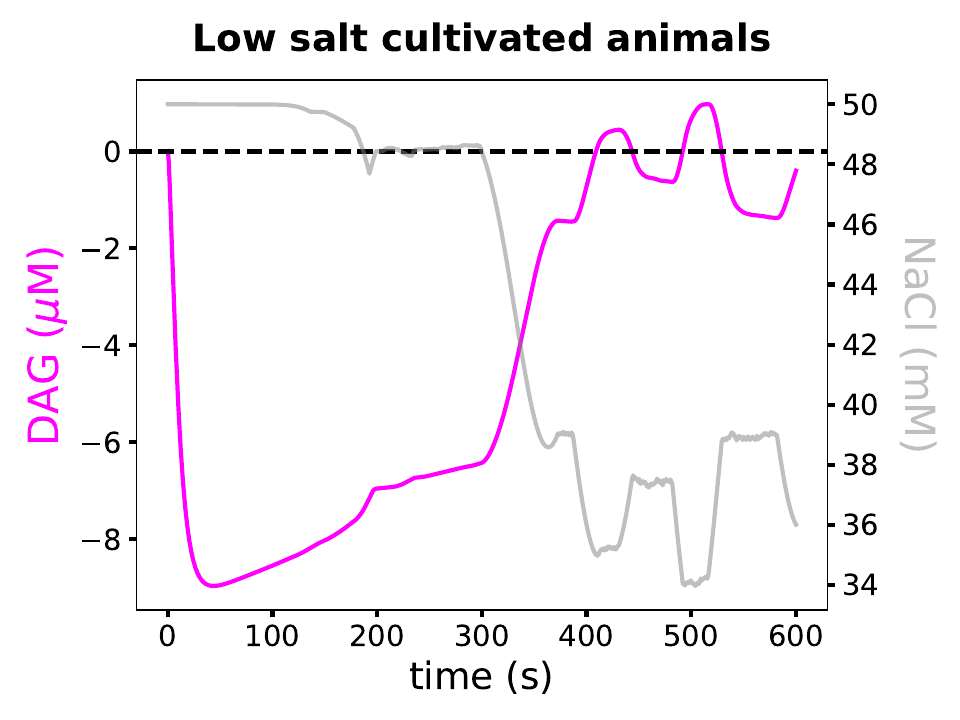}
\end{minipage}
\caption{\textbf{DAG dynamics and perceived NaCl concentration during the chemotaxis assay.} Example time series of DAG levels during the chemotaxis assay. The dashed line shows the threshold $\theta$ of Eq.~\eqref{glutamate}. (a) For high-salt cultivated animals DAG levels increase at the start of the assay due to the sensed decrease of [NaCl]. Increased DAG levels ensure an excitatory synapse between ASER and AIB, which guides the worm to higher [NaCl] areas. (b) For low-salt cultivated animals DAG decreased levels enable an inhibitory synapse between ASER and AIB, which guides the worm towards lower [NaCl] areas. }
\label{dag_assay}  
\end{figure}

\section{Behavioral modeling}
\label{sec:results}

Having studied the response of the different molecular and cellular model variables to NaCl steps, we now aim to see whether this minimal sensory system, alongside with the simplified motor system also described above, can give rise to the behaviors observed experimentally when worms are placed in a continuous gradient, instead of facing sharp NaCl steps.
To that end, we have followed the chemotaxis assay protocol depicted schematically in Fig. \ref{fig:1} and previously described in \cite{kunitomo2013concentration, hiroki2022molecular}.
In the experiment, worms were cultivated at a constant concentration of NaCl (25~mM, 50~mM, or 100~mM) for 6 hours. In those conditions, any perturbations in DAG levels during the pre-cultivation phase fully vanish and DAG activity returns to baseline.
The animals are then transferred to the center of a plate with a salt gradient, and allowed to move freely for about $\sim$10~min.
We have used a virtual gradient similar to the experimental one \cite{kunitomo2013concentration, hiroki2022molecular}, which has a low ($\sim$30~mM) and high ($\sim$95~mM) NaCl area, with a background salt concentration in the rest of the plate of 50~mM, as shown in Fig. \ref{gradiente_plate}.
The simulated gradient is given by the following equation:
\begin{align}
 [\mathrm{NaCl}](x,y) = C_{\mathrm{back}} & + C_{\mathrm{max}}e^{-\frac{(\mathbf{x}
 - \mathbf{x}_{\mathrm{max}})^{2}}{2\sigma ^{2}}} \nonumber \\
 & - C_{\mathrm{min}}e^{-\frac{(\mathbf{x} - \mathbf{x}_{\mathrm{min}})^{2}}{2\sigma ^{2}}}
 \label{eq:gradient}
\end{align}
We also tested our model using other NaCl gradient shapes, such as a unimodal gaussian and a conical function (linear with distance to the peak).
In all cases, we obtained consistent results, which shows our results are not dependent on the particular shape of the gradient.

Figure~\ref{trajectories} shows typical trajectories of ten virtual worms place in the gradient \eqref{eq:gradient}, as generated by our integrated model, with different initial orientations ($\varphi$ in Eq.~\ref{position}), for pre-assay cultivations at 25~mM, 50~mM and 100~mM, respectively.
As shown in the figure, cultivation with different salt concentrations leads to correspondingly different [NaCl] preferences in the chemotaxis assay, as seen experimentally \cite{kunitomo2013concentration}.
The different behavioral preferences of Fig. \ref{trajectories} arise from the same minimal sensorimotor system of Fig. \ref{fig:circuit}, without any change in the model parameters. 

As explained before, the essential mechanism underlying this experience-dependent behavior is that DAG levels in ASER are dynamically regulated according to salt changes.
To see this more clearly, Fig.~\ref{dag_assay} shows DAG levels during the chemotaxis assay for an representative virtual worm.
For high-salt cultivated animals DAG increases at the start of the assay due to the sensed [NaCl] decrease when they are transferred from the cultivation plate (where [NaCl] is 100~mM) to the center of the gradient plate (where [NaCl] is $\sim$50~mM).
This increase in DAG levels above the threshold $\theta$ of Eq.~\eqref{glutamate} (dashed line in Fig.~\ref{dag_assay}a) leads to an excitatory synapse between ASER and AIB, which gives rise to high salt attraction.
As we can see in Fig.~\ref{dag_assay}a, as the worm migrates to higher [NaCl] areas in the plate, DAG levels decrease accordingly.

The situation is reversed in the case of low-salt cultivated animals.
There, DAG decreases at the start of the assay due to the worm being transferred from its cultivation plate (where [NaCl] is 25~mM) to the center of the gradient plate (where [NaCl] is $\sim$50~mM).
Decreased DAG levels lead to an inhibitory synapse between ASER and AIB, which gives rise to low salt attraction.
As we can see in Fig.~\ref{dag_assay}b, DAG levels increase according to migration towards lower [NaCl] areas.

\section{Chemotaxis assay for simulated mutant worms}

To validate quantitatively the model described so far, we now compare the behavior of wild-type worms with mutants where some system components are altered or eliminated altogether.
Some of the mutants studied have been characterized experimentally, and thus are here used as validations of the model, while others have not been studied in the laboratory, and can thus be considered model predictions.

To quantify the behavior of the worms we use the chemotaxis index (C.I.), defined as \cite{kunitomo2013concentration}
\begin{equation}
\mathrm{C.I.} = \frac{N_\mathrm{High}-N_\mathrm{Low}}{N-N_\mathrm{Start}},
\label{eq:CI}
\end{equation}
where $N_\mathrm{High}$ ($N_\mathrm{Low}$) denotes the number of animals in the high (low) NaCl area, i.e. within a 1.05~cm radius from $\mathbf{x}_{\mathrm{max}}$ ($\mathbf{x}_{\mathrm{min}}$) --see Eq.~\eqref{eq:gradient}--, at the end of the assay. $N$ is the total number of simulated worms, and $N_\mathrm{Start}$ is the number of animals within a 1-cm radius from the start point at the end of the assay (the center of the plate).
We consider $N=100$ worms in every assay.

A C.I. of 1 and -1 represents total preferences for high and low concentrations, respectively.
A C.I. of 0, on the other hand, can represent a preference for the background concentration (50~mM), equal distribution of the population between both end areas, or a random distribution.
All the parameters characterizing the NaCl gradient and the C.I. areas are given in Table~\ref{tab:gradient}.

\begin{figure*}[ht]
\centerline{
\includegraphics[width=0.95\textwidth]{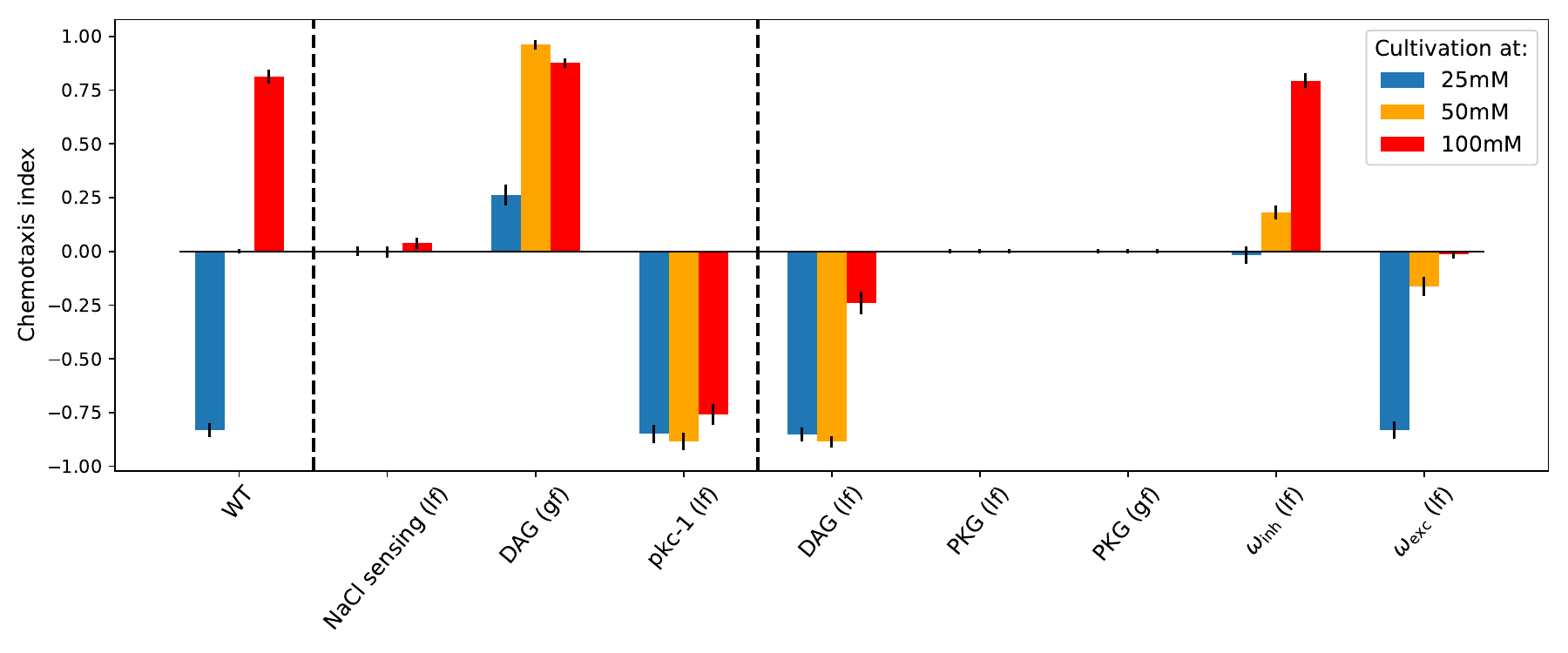}
}
\caption{\textbf{Chemotaxis index for different pre-assay cultivations}. Results for wild-type worms (WT) and eight mutants are displayed. Mean~$\pm$~s.e.m.; $n=6$ assays. 100 worms were simulated in each assay.}
\label{wt_mutant}  
\end{figure*}

Figure~\ref{wt_mutant} shows the chemotaxis indices exhibited by the wild-type (WT) worms and eight mutants for low and high NaCl pre-cultivation concentrations, and for the control case in which the worms are cultivated at the intermediate assay concentration (50~mM). 
In the WT case, the chemotaxis indices confirm the behavior reported in Fig.~\ref{trajectories}: worms cultivated at low [NaCl] (25~mM) are attracted to low salt concentrations (C.I. $\sim$-1), whereas animals cultivated at high [NaCl] (100~mM) are attracted to high salt concentrations (C.I. $\sim$1). Lastly, for worms cultivated at the background concentration of the gradient plate (50~mM), the C.I. $\sim$0, which corresponds in this case to a preference for that same concentration. These chemotaxis indices are in close agreement with the experimental results of Ref. \cite{hiroki2022molecular}. 

Besides the wild-type worms, Fig.~\ref{wt_mutant} shows the chemotaxis indices for two types of mutants.
Specifically, the three cases shown in the middle section of the figure correspond to animals that have been previously studied experimentally in a gradient similar to the one of Fig.~\ref{gradiente_plate}.
Additionally, the five cases at the right of the plot correspond to mutants for which there are not experimental measurements, to our knowledge.
 
The first mutant corresponds to NaCl sensing loss-of-function (lf), which is associated with the inhibition on cGMP production by NaCl.
We have modeled this situation by decreasing the value of the parameter $\alpha$ in Eq.~\eqref{cgmp}.
As can be seen in Fig.~\ref{wt_mutant}, the C.I. for this mutant is close to zero irrespective of the pre-assay cultivation.
This result coincides with the one reported in Ref.~\cite{kunitomo2013concentration}, where the authors examined a mutant strain severely impaired in its ability to sense chemicals (\textit{dyf-11}).

Next we aimed at replicating a gain-of-function (gf) mutation on the DAG/PKC-1 signalling pathway, which corresponds to a higher baseline level of DAG.
To that end, we added in Eq.~\eqref{dag_dyn} a constant production rate $\alpha_\mathrm{DAG}$ (independent of $\mathrm{Ca}^{2+}$ levels):
\begin{equation}
 \frac{d \mathrm{DAG}}{d t} = \alpha_\mathrm{DAG} + \beta_{\mathrm{DAG}}\mathrm{Ca}^{2+} -\delta_{\mathrm{DAG}}\mathrm{DAG}
\label{dag_dyn_mut}
\end{equation}
As shown in Fig.~\ref{wt_mutant} for DAG (gf), these mutant worms migrate towards higher NaCl concentration regardless of the at pre-assay cultivation concentration. 
This can be understood by taking into account that as DAG levels are increased, the worms are mostly in the regime of an excitatory synapse between ASER and AIB for all pre-assay conditions, which explains the resultant overall high salt attraction.
This behavior agrees with experimental results of a gf mutant in the DAG pathway (\textit{egl-30}) \cite{kunitomo2013concentration}.

The DAG/PKC-1 signalling pathway has been further explored experimentally with a PKC-1 loss of function (lf) mutant \cite{hiroki2022molecular}.
PKC-1 is a protein kinase downstream of DAG, which regulates the basal glutamate release in ASER ($\mathrm{Glu_{basal}}$ in our model) by modulating the phosphorylation of a target protein.
We simulated this lf mutant by setting $\alpha_{\mathrm{Glu}}=0.0$ in Eq.~\eqref{glutamate}, which leads to a constitutively low level of $\mathrm{Glu_{basal}}$.
In this case, the model predicts (fourth set of bars in Fig.~\ref{wt_mutant}) that worms with this mutation have a preference for low salt concentrations irrespective of the pre-assay cultivation conditions.
These results are in agreement with the ones reported in Ref.~\cite{hiroki2022molecular} for PKC-1 lf mutants. 

Another way of decreasing the activity of the DAG/PKC-1 signalling pathway is by acting directly on DAG.
We implemented this perturbation by adding a constant degradation rate in Eq.~\eqref{dag_dyn}.
Similarly to the PKC-1 lf mutant (and oppositely to the DAG gf mutant) the DAG baseline is decreased, which leads to a predominantly inhibitory synapse between ASER and AIB.
Consequently, the model predicts again that worms will move toward lower NaCl concentrations for all pre-assay conditions, as shown in Fig.~\ref{wt_mutant}.
This is similar to the behavior of the PKC-1 lf mutant described above, but in this case the behavior is due to low DAG levels, instead of low $\mathrm{Glu_{basal}}$ release from ASER.
Interestingly, in the DAG lf mutant worms cultivated at high salt concentrations (100~mM), the attraction to low salt concentration is weaker than in PKC-1 lf mutants discussed above. 

To explore our hypothesis that NaCl sensory transduction takes place via a feedforward circuit mediated by PKG, we next simulated loss- and gain-of-function PKG mutants.
We did so by setting the value of $\gamma$ in Eq.~\eqref{pkg} to zero or by rising its value, respectively.
In both cases, worms are predicted to move randomly for all cultivation conditions, as shown in Fig. \ref{wt_mutant}.
The fact that these opposite mutants lead to the same behavior can be explained from the symmetric response of the AIB neurons to low and high glutamate levels, as given by the U-shaped response curve discussed in Section \ref{sec:dual} and shown in Fig.~\ref{fig:AIB} above.
Low PKG levels in the lf mutant lead to consistently high $\mathrm{Ca}^{2+}$ levels, which places the operating point of the neuron in the saturating region to the right of the AIB response curve shown in Fig.~\ref{fig:AIB}, for which high AIB activity and consequently high turn probability (and random motion of the worm) irrespective of how sensed NaCl is changing.
In the PKG gf mutants, in turn, $\mathrm{Ca}^{2+}$ levels are consistently low, which places the neuron in the saturating region to the left of the AIB response curve (see again Fig.~\ref{fig:AIB}), for which AIB activity is also high, leading once again to frequent turns and random motion in the NaCl gradient.

Finally, we used the model to probe the relevance of the dual synaptic character in the experience-encoding, by considering loss-of-function mutants that alter the inhibitory and excitatory strengths of glutamate signaling from ASER to AIB. 
Our simulations show that when the inhibitory current in AIB is eliminated by setting $\omega_{\mathrm{inh}}=0.0$ in Eq.~\eqref{vaib}, worms cultivated at low [NaCl] lose their low-salt attraction (next-to-last case in Fig. \ref{wt_mutant}), while high-salt attraction is retained for worms cultivated at high [NaCl]. In turn, when the excitatory current in AIB is eliminated by setting $\omega_{\mathrm{exc}}=0.0$, high-salt cultivated worms do not have a [NaCl] preference, while the behavior of low-salt cultivated worms stays the same as the WT (last case in Fig. \ref{wt_mutant}). All the modified parameter values used for the mutants can be found in Table~\ref{tab:param_mut}. 

\section{Discussion}
\label{sec:discussion} 

In this work we have developed a mechanistic model that captures experience-dependent chemotaxis to NaCl in \textit{C. elegans}. Specifically, we wanted to model how worms migrate to the salt concentration at which they have been previously fed \cite{kunitomo2013concentration}.
Through a set of elegant experiments, Yuichi Iino and collaborators have identified the key molecular and cellular factors associated with this behavior.
In particular, DAG levels in the salt-sensing neuron ASER are regulated in response to the difference between the previously and currently perceived NaCl levels.
DAG activity, in turn, affects the glutamate-modulated connection between ASER and the motor neuron AIB, which can be switched between excitatory and inhibitory.
By this mechanism, animals are guided toward the NaCl concentration they have previously experienced.

Based on these experimental observations, we have built a minimal integrated sensorimotor model of the experience-dependent chemotaxis to NaCl.
The model covers the molecular, cellular and behavioral scales within a parsimonious framework: it is biologically grounded yet not too overly complex.
One of the salient features of our model is that we can explain the two experience-dependent behaviors, high-salt and low-salt attraction, with a single framework without any modification in the model parameters.
Salt preference arises from dynamically regulating DAG levels according to changes in the perceived NaCl concentration.
Besides being able to reproduce the observed chemotaxis behavior of wild-type worms, our minimal model also reproduces the behavior of mutants involving molecular elements of both the sensor and actuator components of the system.
Additionally, we have simulated untested mutations in ASER circuitry and in post-synaptic AIB neuron, whose behavior constitute model predictions to be tested experimentally.

While our work is based on experimental measurements of ASER's calcium and DAG dynamics in response to step changes in NaCl concentration \cite{suzuki2008functional,hiroki2022molecular,ohno2017dynamics}, the results of our model show that the proposed mechanism can also explain the observed behavior in continuous salt gradients. 
The model is also relevant because its essential principle, i.e. the modulation of the DAG signalling pathway, is not specific of chemotaxis to NaCl, but is also involved in experience-dependent behaviors in response to other sensory cues such as temperature (thermotaxis) and volatile odorant (odor chemotaxis), among others \cite{okochi2005diverse, tsunozaki2008behavioral}. 

Finally, our model of NaCl chemotaxis is a suitable starting point to explore the interplay between different stimuli, such NaCl and nutrients.
It has been shown that pairing starvation with exposure to NaCl reverses the well-fed behavior \cite{saeki2001plasticity}: after being starved in the presence of high (low) NaCl, \textit{C. elegans} moves towards low (high) concentrations if set in a gradient \cite{kunitomo2013concentration}. 
DAG dynamics is known to be located downstream of the insulin signaling pathway \cite{ohno2017dynamics, rahmani2021investigating}, which is in turn required for the starvation response \cite{tomioka2006insulin}.
This could explain why the DAG response to NaCl changes is diminished in starved worms in comparison with the well-fed case \cite{ohno2017dynamics}.
Similarly to what we have done in this paper, mathematical modeling could be used to put together these observations to provide us with an integrated perspective on how living organisms respond to multivariate time-dependent signals from their environment.

\section*{Acknowledgements}

We thank Prof Eduardo Izquierdo for useful conversations, and Prof Yuichi Iino and his collaborators for their inspiring and systematic experimental investigations of chemotaxis in \textit{C. elegans}. 
This work was supported by project PID2021-127311NB-I00 financed by the Spanish Ministry of Science and Innovation, the Spanish State Research Agency and FEDER
(MICIN/AEI/10.13039/ 501100011033/FEDER), by the Maria de Maeztu Programme for Units of Excellence in R\&D (project CEX2018-000792-M), by the ICREA Academia programme, and by the Fundación Tatiana Pérez de Guzmán el Bueno.
M.S.V. is supported by a PhD fellowship from the Ag\`encia de Gesti\'o d'Ajuts Universitaris i de Recerca (AGAUR) from the Generalitat de Catalunya (grant 2021-FI-B-00408).


%

%

%

\clearpage

\section*{Tables}


\begin{table}[h]
\begin{tabular}{|c|c|c|}
\hline
 Parameter & Value & Equation \\ \hline
 $\alpha$ & 825~$\mu\mathrm{M}\cdot\mathrm{s}^{-1}$ & \eqref{cgmp} \\ \hline
 $K$ & 300 mM & \eqref{cgmp} \\ \hline
 $\delta_{\mathrm{GMP}}$ & 50 $\mathrm{s}^{-1}$ & \eqref{cgmp} \\ \hline
 $\gamma$ & $0.12$ & \eqref{pkg} \\ \hline
 $\delta_{\mathrm{PKG}}$ & 0.12~$\mathrm{s}^{-1}$ & \eqref{pkg} \\ \hline
 $\beta$ & $1.0\,\mu\mathrm{M}\cdot\mathrm{s}^{-1}$ & \eqref{calcium} \\ \hline
 $\delta_{\mathrm{Ca}}$ & 1.0 $\mathrm{s}^{-1}$ & \eqref{calcium} \\ \hline
 $b$ & 2.0 $\mu$M$^{-1}$ & \eqref{tanh} \\ \hline
 $\beta_{\mathrm{DAG}}$ & 0.7~$\mathrm{s}^{-1}$ & \eqref{dag_dyn} \\ \hline
 $\delta_{\mathrm{DAG}}$ & 0.001~$\mathrm{s}^{-1}$ & \eqref{dag_dyn} \\ \hline
 $\alpha_{\mathrm{Glu}}$ & 1.345~mM & \eqref{glutamate} \\ \hline
 $\beta_{\mathrm{Glu}}$ & 0.055~mM & \eqref{glutamate} \\ \hline
 $\alpha_{\Delta}$ & $1000$ & \eqref{glutamate} \\ \hline
 $\theta$ & 0~$\mu$M & \eqref{glutamate} \\ \hline
 $\tau$ & 0.1 s & \eqref{vaib} \\ \hline
 $ \omega_{\mathrm{inh}}$ & 10.0 mV & \eqref{vaib} \\ \hline
 $\omega_{\mathrm{exc}}$ & 50.0 mV & \eqref{vaib} \\ \hline
 $V_{\mathrm{rest}}$ & -55.0 mV & \eqref{vaib} \\ \hline
 
 $b_{\mathrm{exc}}$ & 27~mM$^{-1}$ & \eqref{s_exc} \\ \hline
 $\theta_{\mathrm{exc}}$ & 1.481~mM & \eqref{s_exc} \\ \hline
  $b_{\mathrm{inh}}$ & 92~mM$^{-1}$ & \eqref{s_inh} \\ \hline
 $\theta_{\mathrm{inh}}$ & 0.054~mM & \eqref{s_inh} \\ \hline
 $\omega_{\mathrm{low}}$ & 0.03 $\mathrm{s}^{-1}$ & \eqref{pirouette} \\ \hline
 $\omega_{\mathrm{high}}$ & 50.3 $\mathrm{s}^{-1}$ & \eqref{pirouette} \\ \hline
 $V_{\mathrm{low}}$ & -50.035 mV & \eqref{pirouette} \\ \hline
 $v$ & 0.022 cm/s  & \eqref{position} \\ \hline
\end{tabular}
\caption{Parameter values of the model}
\label{tab:param}
\end{table}

\begin{table}[h]
\begin{tabular}{|c|c|}
\hline
 Parameter & Value \\ \hline
 $C_{\mathrm{max}}$ & 45.0~mM \\ \hline
 $C_{\mathrm{min}}$ & 20.0~mM \\ \hline
 $C_{\mathrm{back}}$ & 50.0~mM \\ \hline
 $\mathbf{x}_{\mathrm{max}}$ & (3.0, 0.0)~cm \\ \hline
 $\mathbf{x}_{\mathrm{min}}$ & (-3.0, 0.0)~cm \\ \hline
 $\sigma$ & 0.7~cm \\ \hline
 $R$ & 4.25~cm \\ \hline
\end{tabular}
\caption{Parameter values of the gradient \eqref{eq:gradient}}
\label{tab:gradient}
\end{table}

\begin{table}[h]
\begin{tabular}{|l|c|c|c|}
\hline
 Mutant & Parameter  & New value & Equation \\ 
 \hline
NaCl (lf) & $\alpha$ & 0.0825~$\mu\mathrm{M} \cdot\mathrm{s}^{-1}$ & \eqref{cgmp} \\ \hline
DAG (gf) & $\alpha_\mathrm{DAG}$ & 0.01~$\mu\mathrm{M}\cdot\mathrm{s}^{-1}$ & \eqref{dag_dyn_mut} \\ \hline
pkc-1 (lf) & $\alpha_{\mathrm{Glu}}$ & 0.0 mM & \eqref{glutamate} \\ \hline
DAG (lf) & $\alpha_\mathrm{DAG}$ & -0.01~$\mu\mathrm{M}\cdot\mathrm{s}^{-1}$ & \eqref{dag_dyn_mut} \\ \hline
PKG (lf) & $\gamma$ & 0.0 & \eqref{pkg} \\ \hline
PKG (gf) & $\gamma$ & 1.0 & \eqref{pkg} \\ \hline
$\omega_{\mathrm{inh}}$ (lf) & $ \omega_{\mathrm{inh}}$ & 0.0 mV & \eqref{vaib} \\ \hline
$\omega_{\mathrm{exc}}$ (lf) & $ \omega_{\mathrm{inh}}$ & 0.0 mV & \eqref{vaib} \\ \hline
\end{tabular}
\caption{Parameter values for the mutants}
\label{tab:param_mut}
\end{table}

\end{document}